\documentclass[useAMS,usenatbib]{mn2e}
\usepackage[dvips]{graphicx}
\usepackage{textcomp}
\usepackage{times}
\usepackage{float,lscape}
\usepackage{subfig}
\usepackage{amsmath}
\usepackage{amssymb}
\def\ks{{$K_{\rm s}$}}
\def\FoVH{{10$'\times$10$'$}}
\def\FoVT{{5$'\times$5$'$}}

\usepackage[T1]{fontenc}
\title[Sh2-138: a small cluster of massive stars]{Sh2-138: Physical environment around a small cluster of massive stars}
\author[T. Baug et al.]{T. Baug,$^1$\thanks{E-mail: tapas.baug@tifr.res.in (TB)} D.K. Ojha,$^1$ L.K. Dewangan,$^2$ J.P. Ninan,$^1$ B.C. Bhatt,$^3$ S.K. Ghosh $^{1,4}$ and \newauthor K.K. Mallick$^1$\thanks{This file has been amended to highlight the proper use of \LaTeXe\ code with the class file.}\\$^1$Tata Institute of Fundamental Research, Homi Bhabha Road, Colaba, Mumbai 400005, India\\$^2$Instituto Nacional de Astrof\'{i}sica, \'{O}ptica y Electr\'{o}nica, Luis Enrique Erro \# 1, Tonantzintla, Puebla, M\'{e}xico C.P. 72840, M\'{e}xico \\$^3$Indian Institute of Astrophysics, Koramangala, Bangalore 560 034, India\\$^4$National Centre for Radio Astrophysics, Tata Institute of Fundamental Research, Pune 411 007, India}
\begin{document}

\date{Submitted}

\pagerange{\pageref{firstpage}--\pageref{lastpage}} \pubyear{2015}

\maketitle

\label{firstpage}
\begin{abstract}
We present a multi-wavelength study of the Sh2-138, a Galactic compact H {\sc ii} region. The data comprise of optical and near-infrared (NIR) photometric and spectroscopic observations
 from the 2-m Himalayan {\it Chandra} Telescope, radio observations from the Giant Metrewave Radio Telescope (GMRT), and archival data covering radio through NIR wavelengths. A total of 10 Class I and 54 Class II
 young stellar objects (YSOs) are identified in a 4$\farcm$6$\times$4$\farcm$6 area of the Sh2-138 region. Five compact ionized clumps, with four lacking of any optical or NIR counterparts, are identified using the
 1280 MHz radio map, and correspond to sources with spectral type earlier than B0.5. Free-free emission spectral energy distribution fitting of the central compact H {\sc ii} region yields an electron density of
 $\sim$2250$\pm$400 cm$^{-3}$. With the aid of a wide range of spectra, from 0.5-15 $\mu m$, the central brightest source - previously hypothesised to be the main ionizing source - is characterized as a Herbig Be type star.
 At large scale (15$'$ $\times$15$'$), the {\it Herschel} images (70--500 $\mu m$) and the nearest neighbour analysis of YSOs suggest the formation of an isolated cluster at the junction of filaments. Furthermore, using
 a greybody fit to the dust spectrum, the cluster is found to be associated with the highest column density ($\sim$3$\times$10$^{22}$ cm$^{-2}$) and high temperature ($\sim$35 $K$) regime, as well as with the radio
 continuum emission. The mass of the central clump seen in the column density map is estimated to be $\sim$3770 $M_\odot$.
\end{abstract}


\section{Introduction}
Massive stars ($>$ 8 M$_\odot$) influence the evolution of their host galaxies in a multitude of ways, through stellar winds, outflows, expanding H {\sc ii} regions, and supernova explosions \citep{zinnecker07}.
 They are also the primary source of heavy elements, and their ultraviolet (UV) radiation can inject vast amounts of energy and momentum into the natal medium. However, the formation and interaction of massive stars
 with their surrounding environment is not yet well understood, though it is well-accepted that they are usually associated with clusters \citep{duchene13}. This empirical property of massive stars requires thorough
 observational studies of young stellar clusters in the Galaxy. More recently, with the advent of far-infrared (FIR) and submillimetre (sub-mm) observations, such clusters have been often found to be associated with
 filamentary structures \citep{andre10,schneider12}. However, a study of such young stellar clusters, along with the role of filaments in the formation and evolution of dense massive star-forming clumps, is currently
 a matter of active investigation.
 
Sh2-138 \citep{sharpless59} is a Galactic compact H {\sc ii} region ($\alpha_{2000} \sim$ 22$^{\rm h}$32$^{\rm m}$46$^{\rm s}$, $\delta_{2000}\sim$ +58$^{\rm d}$28$^{\rm m}$22$^{\rm s}$), associated with IRAS 22308+5812
 (also referred as $IRAS$ source in this work).
 It harbors a stellar cluster which is dominated by at least four O--B2 stars \citep{deharveng99}, with a total luminosity of $\sim$4.9$\times$10$^4$ L$_\odot$ \citep{simpson90}. \citet{deharveng99} studied the optical
 spectra of the brightest object present in the stellar cluster and suggested that this source could be a Herbig Ae/Be candidate. Radio continuum emission at 4.89 GHz (beam $\sim$13$\arcsec$), with a diameter of
 $\sim$1$\farcm$5 ($\sim$2.5 pc at a distance of 5.7 kpc), was detected near the $IRAS$ source by \citet{fich93}. A study of the densest parts of this region, as well as the molecular boundaries, was carried out by
 \citet{johansson94} using the isotopomers of CO, CS, SO, CN, HCN, HNC, HCO$^{+}$ and H$_{2}$CO. \citet{qin08} reported bipolar molecular outflows towards the $IRAS$ source with an entrainment rate of $\sim$22.5
 $\times$ 10$^{-4} M_\odot$ ${\rm yr ^{-1}}$  using the CO $J=3-2$ and $J=2-1$ line emissions. Electron density calculation, based on the fine structure lines of [Ar {\sc ii}--{\sc iii}], [S {\sc iii}], and [Ne {\sc ii}]
 from the Infrared Space Observatory (ISO) spectra, suggests that the Sh2-138 is a classical H II region \citep{martin02}.

\citet{deharveng99} use a distance estimate of 5 kpc for the Sh2-138 region, assuming it to be a part of the same complex as the nearby regions with similar velocities. They basically use the mean distance of
 these other nearby regions, whose individual distances have large variation, from 3.45-5.9\,kpc. However, the kinematic distance calculation by \citet{deharveng99}, based on the CO ($J=1-0$) observations of
 \citet{blitz82}, yields a distance of 5.9$\pm$1.0\,kpc. A similar distance of $\sim$5.7 kpc is calculated by \citet{wouterloot89} using radial velocity data from their $^{12}$CO ($J=1-0$) observations. Hence, we
 adopt a value of 5.7$\pm$1.0 kpc for our work.

The previous studies on the Sh2-138 region using optical and near-infrared (NIR) observations have demonstrated the presence of ongoing star formation in a stellar cluster containing at least four massive stars. However,
 those studies were mainly focused within an area of $\sim$2$\arcmin\times$2$\arcmin$ encompassing the stellar cluster. An overall morphological study of the region, and its relation to the ongoing star formation and stellar
 population, is still pending. Additionally, the study of physical environment of the Sh2-138 region at a large scale is yet to be explored observationally. Furthermore, the advent of FIR and sub-mm observatories like
 {\it Herschel} has provided opportunity to explore the large-scale structures, as well as investigate theories such as the role of filaments in the formation of stellar clusters \citep{myers09,schneider12,mallick13,mallick15}.
 Since star formation in a region is a sum of many components, it is helpful to carry out an overall study at various wavelengths. In this work, we have performed a detailed multi-wavelength analysis of the Sh2-138 region
 from 25 pc to 0.1 pc scale centered on IRAS 22308+5812. This has been done using new optical and NIR photometric and spectroscopic observations, as well as radio observations, from Indian observational facilities, complemented
 with the multi-wavelength data covering radio through NIR wavelengths from publicly available surveys.

In Section 2, we present the observations of the Sh2-138 region and data reduction techniques. Other available archival data sets used in this paper are summarized in Section 3. Morphology of the region inferred using
 multi-wavelength observations is presented in Section 4. In Section 5, we discuss the nature of the brightest star detected in the region using optical, NIR, and mid-infrared (MIR) spectra. The identification and
 selection of young stellar objects (YSOs) are presented in Section 6, followed by a general discussion in Section 7, and a presentation of the main conclusions in Section 8.
\section{Observations and data reductions}
\subsection{Optical $BVRI$-band Photometry}
Optical $BVRI$ imaging observations of the Sh2-138 region were carried out on 2005 September 8 using the Himalaya Faint Object Spectrograph and Camera (HFOSC) mounted on the 2\,m Himalayan $Chandra$ Telescope (HCT).
 HFOSC is equipped with a SITe 2k$\times$4k pixels CCD and the central 2k$\times$2k pixels of the CCD are used for imaging. With a plate scale of $\sim$0.3 arcsec pixel$^{-1}$, it covers a field of view of
 $\sim$10$'\times$10$'$ on the sky. Images were obtained with long and short exposures in the Bessell $B$ (600s, 60s, 20s), $V$ (600s, 20s, 5s), $R$ (300s, 20s, 5s), and $I$ (200s, 10s, 3s) band filters. Bias and twilight
 flat frames were also observed in each filter. The $BVRI$ images of the standard field SA 114-176 \citep{landolt92} were obtained for photometric calibration as well as for the extinction coefficients estimation.
 Observations were performed with an average seeing of 1$\farcs$5 full width at half maximum (FWHM).
 
Basic processing of the CCD frames was done using the {\sc iraf}\footnote[1]{Image Reduction and Analysis Facility (IRAF) is distributed by the National Optical Astronomy Observatory, USA} data reduction package. The
 astrometric calibration of these frames was performed using the Two Micron All Sky Survey \citep[2MASS;][]{skrutskie06} coordinates of 12 point sources (spread throughout the frame), and a positional accuracy better than
 $\pm$0$\farcs$08 was obtained. Due to the crowded nature of the Sh2-138 region, point spread function (PSF) photometry was carried out using the IRAF {\sc daophot} package \citep{stetson87}. The PSF was generated from several
 isolated stars ($>$ 9) present in the frame. Magnitudes estimated from both short and long exposure frames, for each filter, were averaged to obtain the instrumental magnitudes. However, we have used magnitudes obtained in
 the short exposure frames for bright sources which are saturated in the long exposure frames. Instrumental magnitudes of broad-band images were converted to the standard values using the colour correction equations available
 for the HFOSC. The 10$\sigma$ limiting magnitudes were found to be 21.3, 22.4, 21.8 and 20.6 for the $B$-, $V$-, $R$-, and $I$-bands, respectively. In $\sim$10$'\times$10$'$ FoV centred on $IRAS$ source, we found a total of
 685, 1979, 2573 and 2582 sources upto the 10$\sigma$ detection limit in $B$-, $V$-, $R$-, and $I$-bands, respectively.
\subsection{Optical Spectroscopy}
\subsubsection{Slitless $H\alpha$ Spectroscopy}
\label{sec_ha_spec}
In order to identify strong $H\alpha$ emission sources in the Sh2-138 region, slitless $H\alpha$ spectra were obtained using the HFOSC on 2007 November 16. The spectra were observed using a combination of grism 5
 (5200-10300 \AA) and wide-$H\alpha$ filter (6300-6740 \AA), with a spectral resolution of 870. The central 2k$\times$2k part of the 2k$\times$4k CCD was utilized for observations. Three slitless spectra, with an exposure
 time of 420s each, were obtained for the region and were finally coadded to increase the signal-to-noise ratio. Image frames without grism were also observed to positionally match the stars with their slitless spectra.
 These observations directly allow us to trace the stars with enhanced $H\alpha$ emission with respect to the continuum.
\subsubsection{Slit Spectroscopy}
Optical spectroscopic observations of the central brightest source were performed using the HFOSC on 2014 November 18. The slit (1$\farcs$92 $\times$ 11$\arcmin$) spectrum was obtained using grism 8 (5800-8350 \AA;
 R$\sim$2190), for an exposure time of 40 min. Several bias frames and FeNe calibration lamp spectrum were also obtained for bias subtraction and wavelength calibration, respectively, of the observed spectrum.
 A spectroscopic standard (HIP 14431) was observed for telluric corrections. The observed spectrum was reduced using the {\sc IRAF} data reduction package.
\subsection{NIR $JH$\ks-band photometry}
\label{sec_nir_phot}
The newly installed TIFR Near Infrared Spectrometer and Imager Camera (TIRSPEC) on the HCT was used for NIR observations on 2014 November 18 under photometric conditions with an average seeing of 1$\farcs$4.
 TIRSPEC is equipped with a 1024$\times$1024 pixels HAWAII-1 PACE array which, with a pixel scale of about 0$\farcs$3, translates to a field of view $\sim$5$\arcmin\times$5$\arcmin$. The details of the instrument
 can be found in \citet{ninan14}. We performed deep photometric observations of the Sh2-138 region in $J$ (1.25 $\mu m$), $H$ (1.65 $\mu m$), and \ks~(2.16 $\mu m$) bands. Multiple frames, with a single frame's
 exposure time being 20s, were acquired at 7 dithered positions. After rejecting bad frames, we obtained 57, 54, and 56 good frames which provide an effective on-source integration time of 1140s, 1080s, and
 1120s in $J$-, $H$-, and \ks-bands, respectively. Flat field and sky frames were also observed for flat correction and sky subtraction of the object frames. The astrometric calibration was done using 2MASS
 coordinates of 30 point sources present in the frames. The positional accuracy was found to be better than $\pm$0$\farcs$15.

A semi-automated script written in PyRAF \citep{ninan14} was used for reduction of the observed images following the standard procedure. Due to crowded field of the region, PSF photometry was performed on NIR images using the
 `ALLSTAR' algorithm of the {\sc daophot} package. Isolated bright stars ($>$9) were used to determine the PSF. The instrumental magnitudes were further corrected using the colour correction
 equations for TIRSPEC \citep[given in][]{ninan14}. The resultant magnitudes were calibrated to 2MASS system using about 15 isolated stars having error $<$0.04 mag. Our final catalog is comprised of sources
 in an area of 4$\farcm$6$\times$4$\farcm$6 after removing stars at the edges of the observed frames. The 10$\sigma$ limiting magnitudes were found to be 18.2, 18.0, and 17.8 for the $J$-, $H$-, and \ks-bands, respectively.
 We found a total of 617, 674, and 703 sources upto the 10$\sigma$ detection limit in $J$-, $H$-, and \ks-bands, respectively.

We have evaluated the completeness limit of TIRSPEC \ks-band image using artificial star experiments. Artificial stars with different magnitudes were added in the image, and then it was determined what fraction
of these added stars were detected, per 0.5 magnitude bin, using {\sc daofind} task in {\sc iraf}. Finally, it was found that the recovery rate was more than 90\% for the sources with \ks~$\leq$14.5 mag.
\begin{table}
\centering
\begin{minipage}{100mm}
\caption{Log of Optical and NIR observations}
\begin{tabular}{@{}lccl@{}}
\hline
Filter            &         Exposure Time (sec)             & Date of      & Instrument/FoV \\
                  &       $\times$ No. of frames            & Obs.         &                \\ \hline
\multicolumn{4}{c}{Imaging}                                                                 \\ \hline
  $B$             & 600$\times$1, 60$\times$1, 20$\times$1  & 2005 Sep 08  & HFOSC/\FoVH    \\
  $V$             & 600$\times$1, 20$\times$1, 5$\times$1   & 2005 Sep 08  & HFOSC/\FoVH    \\
  $R$             & 300$\times$1, 20$\times$1, 5$\times$1   & 2005 Sep 08  & HFOSC/\FoVH    \\
  $I$             & 200$\times$1, 10$\times$1, 3$\times$1   & 2005 Sep 08  & HFOSC/\FoVH    \\
$H\alpha$         & 600$\times$1, 250$\times$1, 50$\times$1 & 2005 Sep 08  & HFOSC/\FoVH    \\
\multicolumn{4}{c}{}                                                                        \\
  $J$             & 20$\times$57                            & 2014 Nov 18  & TIRSPEC/\FoVT  \\
  $H$             & 20$\times$54                            & 2014 Nov 18  & TIRSPEC/\FoVT  \\
  \ks             & 20$\times$56                            & 2014 Nov 18  & TIRSPEC/\FoVT  \\
\multicolumn{4}{c}{}                                                                        \\
Methane off       & 30$\times$17                            & 2014 Jun 06  & TIRSPEC/\FoVT  \\
$[Fe$ {\sc ii}$]$ & 30$\times$15                            & 2014 Jun 06  & TIRSPEC/\FoVT  \\
  $H_2$           & 30$\times$15                            & 2014 Jun 06  & TIRSPEC/\FoVT  \\
$Br\gamma$        & 30$\times$15                            & 2014 Jun 06  & TIRSPEC/\FoVT  \\
 $K$-cont         & 30$\times$15                            & 2014 Jun 06  & TIRSPEC/\FoVT  \\ \hline
\multicolumn{4}{c}{Spectroscopy}                                                            \\ \hline
Slitless-$H\alpha$& 420$\times$3                            & 2007 Nov 16  & HFOSC/\FoVH    \\
\multicolumn{4}{c}{}                                                                        \\
Slit-optical      & 2400$\times$1                           & 2014 Nov 18  & HFOSC/--       \\
\multicolumn{4}{c}{}                                                                        \\
Slit-NIR-$Y$      & 100$\times$8                            & 2014 May 29  & TIRSPEC/--     \\
Slit-NIR-$J$      & 100$\times$8                            & 2014 May 29  & TIRSPEC/--     \\
Slit-NIR-$H$      & 100$\times$8                            & 2014 May 29  & TIRSPEC/--     \\
Slit-NIR-$K$      & 100$\times$8                            & 2014 May 29  & TIRSPEC/--     \\ \hline                 
\end{tabular}
\label{table1}
\end{minipage}
\end{table}
\subsection{NIR spectroscopy}
We obtained NIR spectra of the central brightest source on 2014 May 29, using the TIRSPEC, in NIR $Y$ (1.02-1.20 $\mu m$), $J$ (1.21-1.48 $\mu m$), $H$ (1.49-1.78 $\mu m$), and $K$ (2.04-2.35 $\mu m$) bands. Average
 spectral resolution of TIRSPEC is $\sim$1200. We obtained a total of 8 spectra at two dithered positions in each band, with an exposure time of 100s for each spectrum, which gives on-source integration time of 800s
 in each band. Corresponding NIR continuum and Argon lamp spectra were obtained for continuum subtraction and wavelength calibration, respectively, of the observed spectra. Spectra of a separate spectroscopic standard
 star (HIP 14431) were also observed for telluric correction. All the spectra were reduced using a semi-automated script written in PyRAF \citep{ninan14}. The spectra were extracted using the {\sc apall} task in
 {\sc iraf}. Finally, all the $JHK$-band spectra were flux calibrated using the magnitudes derived from the TIRSPEC photometry. $Y$-band spectrum was calibrated using the flux determined at 1.05 $\mu m$ by interpolating
 $I$- and $J$-band fluxes.
\subsection{Optical and NIR narrow-band imaging}
We conducted optical narrow-band imaging observations of the region in $H\alpha$ filter ($\lambda$ $\sim$6563 \AA, $\Delta\lambda$ $\sim$100 \AA) with exposure times of 600s, 250s, and 50s on 2005 September 8
 using the HFOSC. There was no separate narrow-band continuum image observed, and hence, the optical $R$-band image was used for continuum subtraction.
 
NIR narrow-band observations were also obtained in $[Fe$ {\sc ii}$]$ (1.645 $\mu m$; Bandwidth: 1.6\%), $H_2$ (2.122 $\mu m$; Bandwidth: 2.0\%), and $Br~\gamma$ (2.166 $\mu m$; Bandwidth: 0.98\%) filters on 2014
 June 6 using TIRSPEC mounted on the HCT. Observations were also carried out in Methane off band (1.584 $\mu m$; Bandwidth: 3.6\%) for continuum subtraction of $[Fe$ {\sc ii}$]$ images, and in $K$-continuum band
 (2.273 $\mu m$; Bandwidth: 1.73\%) for continuum subtraction of $H_2$ and $Br\gamma$ images. To obtain continuum-subtracted images, each set of narrow band frames was aligned and transformed to same PSF using
 {\sc IRAF} tasks. The final continuum subtracted $H_2$ and $[Fe$ {\sc ii}$]$ images were binned in 6$\times$6 pixels to enhance the signal-to-ratio of the images.

The log of all optical and NIR observations is given in Table \ref{table1}. 
\subsection{Radio Continuum Observations}
Radio continuum observations of the Sh2-138 region at 610 and 1280 MHz bands were carried out using the Giant Metrewave Radio Telescope (GMRT) on 2002 May 19 (Project Code 01SKG01) and 2002 September 27 (Project
 Code 02SKG01), respectively. The GMRT consists of 30 antennas and each antenna is 45m in diameter. For an approximate `Y'-shaped configuration, 12 antennas of the GMRT are randomly distributed in the central
 1x1 km$^2$ region, and remaining 18 are placed along three arms (arm length upto $\sim$14 km). The maximum baseline of GMRT is about 25 km. Primary beam size is about 43 arcmin and 26 arcmin for 610 and 1280 MHz,
 respectively. More details on the GMRT array can be found in \citet{swarup91}.

Total observations time at 610 and 1280 MHz bands was 2.2 hours and 3.6 hours, respectively. The data reduction was carried out using the Astronomical Image Processing Software ({\sc AIPS}) package, following similar
 procedure as described in \citet{mallick13}. The data were edited to flag out the bad baselines or bad time ranges using the {\sc AIPS} tasks. Multiple iterations of flagging and calibration were done to improve the data
 quality, which was finally Fourier-inverted to make the radio maps. A few iterations of (phase) self-calibration were carried out to remove the ionospheric phase distortion effects. The final 610 and 1280 MHz images
 have a synthesised beamsize of 5$\farcs$5$\times$4$\farcs$8 and 3$\farcs$4$\times$2$\farcs$3, respectively.
 
Our source is located towards the Galactic plane while the corresponding flux calibrator (3C48) is situated away from the Galactic plane. At meter wavelengths, a large amount of radiation comes from the Galactic
 plane which increases the effective antennae temperature. Hence, it was important to correct both 610 and 1280 MHz images for system temperature \citep[see][]{omar02,mallick12,vig14}. It was done by rescaling both the
 images by a correction factor of (T$_{freq}$ + T$_{sys}$)/T$_{sys}$, where T$_{sys}$ is the system temperature obtained from GMRT Web site, and T$_{freq}$ is the sky temperature at the observed frequency (i.e., 610 and
 1280 MHz) towards the source obtained using the interpolated value from the sky temperature map of \citet{haslam82} at 408 MHz.
\section{Archival data}
We have also obtained publicly available multi-wavelength archival data to investigate the ongoing physical processes in the Sh2-138 region.
\begin{figure} 
\begin{tabular}{c}
\includegraphics[width=0.45\textwidth]{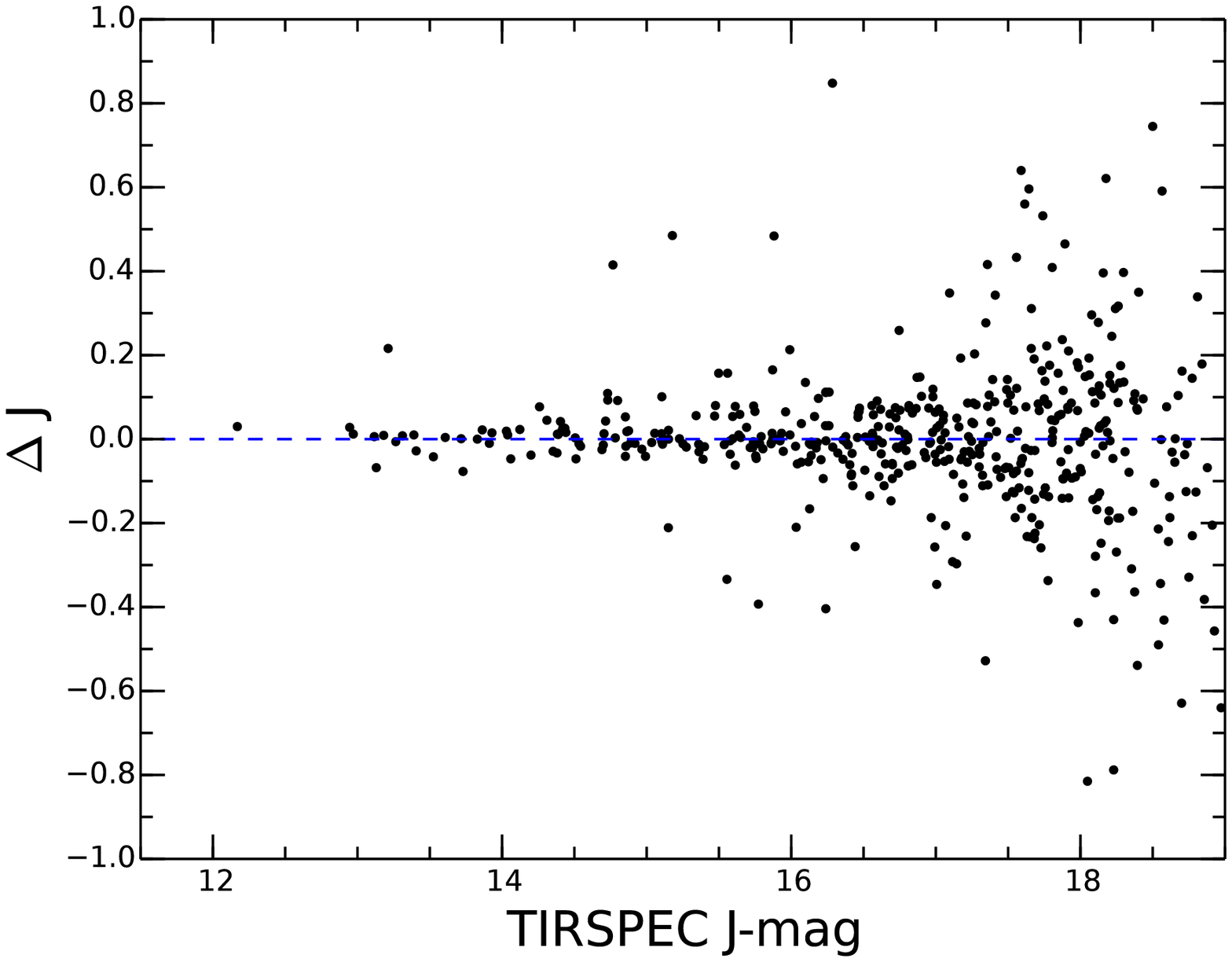} \\
\includegraphics[width=0.45\textwidth]{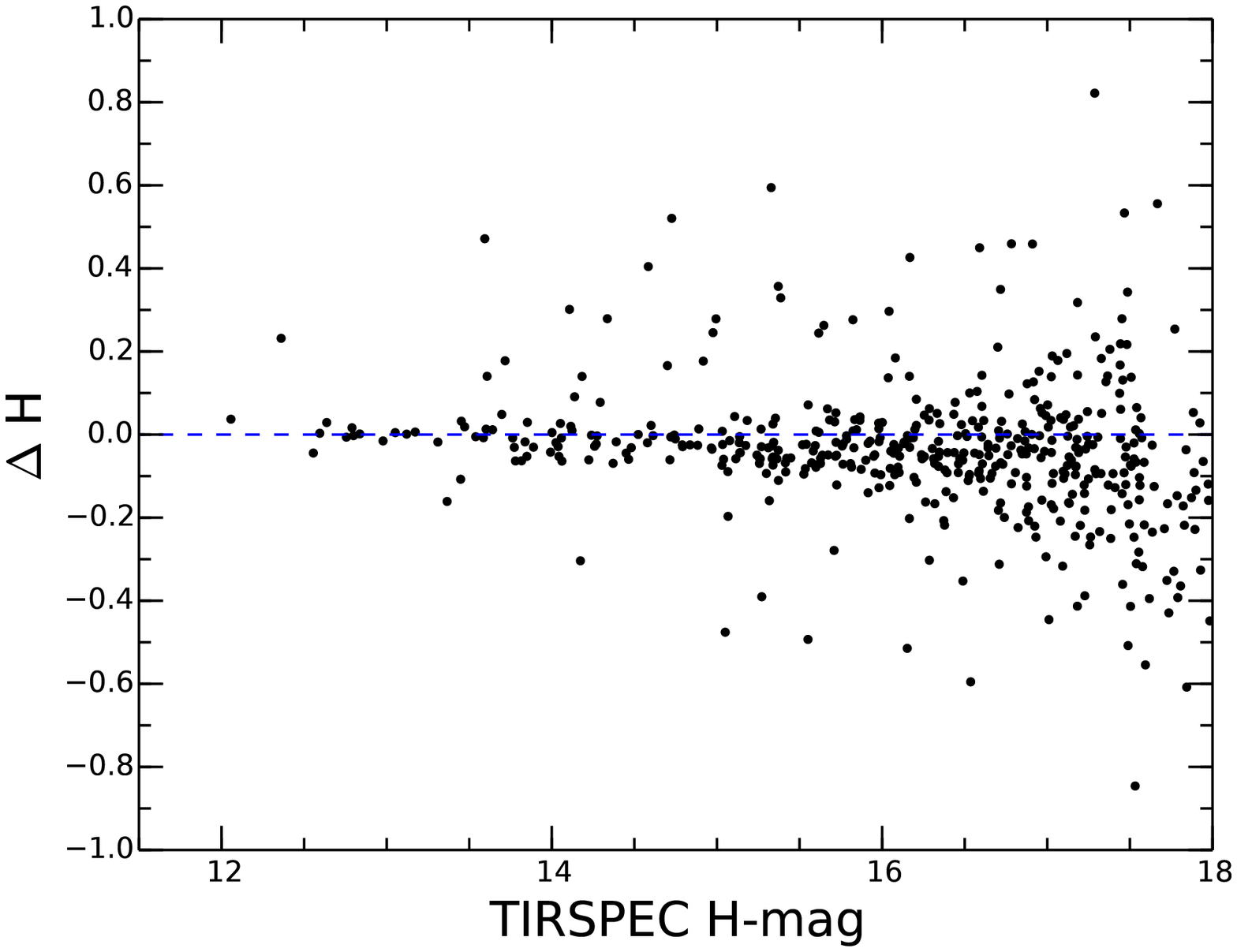} \\
\includegraphics[width=0.45\textwidth]{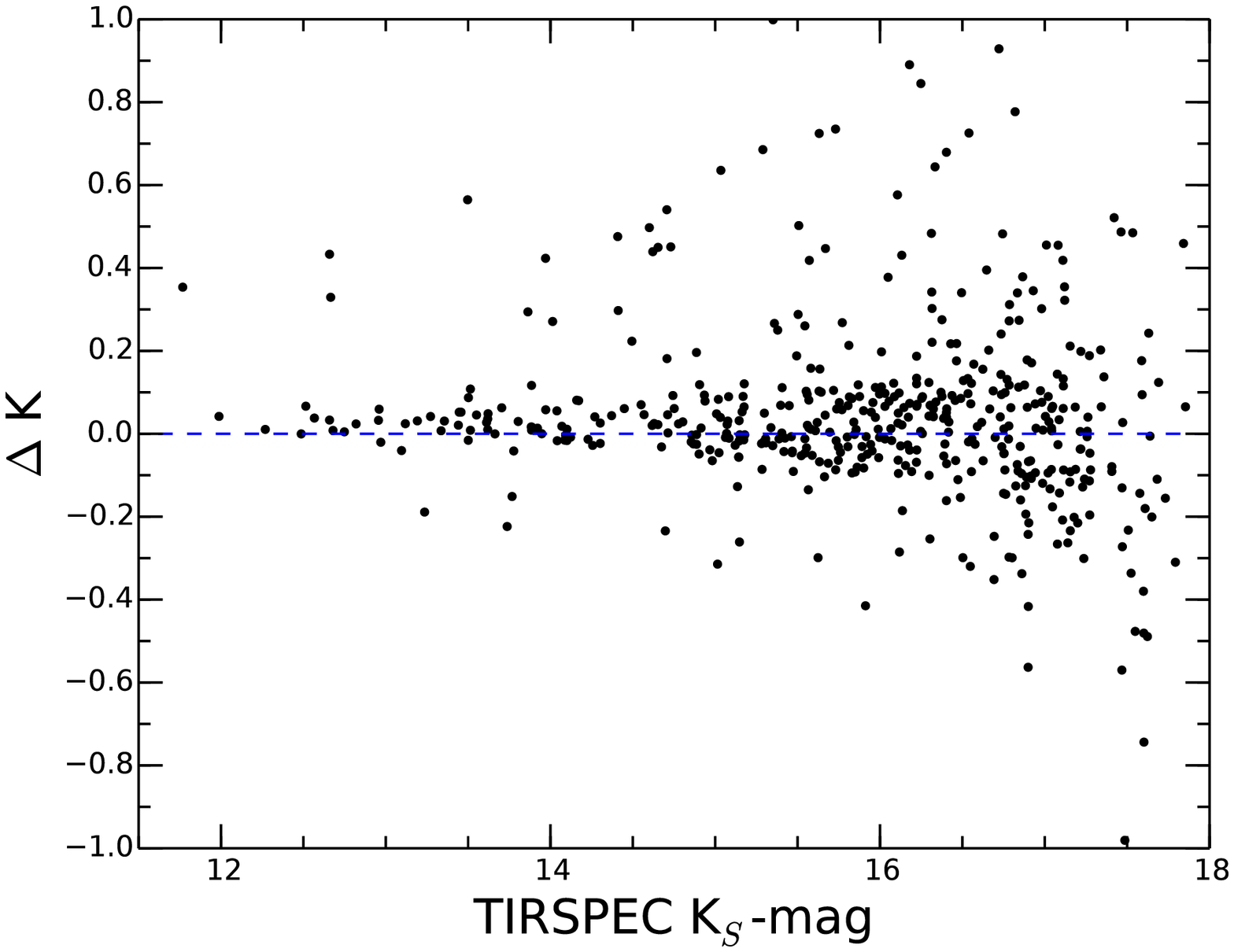} 
\end{tabular}
\caption{A comparison of TIRSPEC and UKIDSS magnitudes. Abacissae in all three figures are TIRSPEC magnitudes and ordinates are differences between UKIDSS and TIRSPEC magnitudes of the respective filter.}
\label{fig0}
\end{figure}
\subsection{UKIDSS {\it JHK} data}
United Kingdom Infrared Deep Sky Survey (UKIDSS) archival data of the Galactic Plane Survey \citep[GPS release 6.0;][]{lawrence07} are available for the Sh2-138 region. UKIDSS observations were obtained using the
 UKIRT Wide Field Camera \citep[WFCAM;][]{casali07}. UKIDSS $J$-band magnitudes are not available for many sources in the central 2$\arcmin\times$2$\arcmin$ nebular region. Therefore, we retrieved the UKIDSS sources
 detected in all the three NIR ($JHK$) bands as well as detected only in the $H$- and $K$-bands. Note that UKIDSS (with 3.8\,m telescope) data allow us to identify much redder sources than TIRSPEC (with 2\,m telescope) in
 $H$- and \ks-bands. For example, sources are detected upto $H-K~\sim~$1.6 using TIRSPEC, while the reddest source detected in UKIDSS in the same field of view (FoV) has $H-K~\sim~$3.1. We selected only reliable photometry
 following the conditions given in \citet{lucas08}. The TIRSPEC and UKIDSS catalogs were cross-matched in the central 4$\farcm$6$\times$4$\farcm$6 area with a matching radius of 1$\arcsec$. The redder UKIDSS sources,
 which are not detected in TIRSPEC, were combined with TIRSPEC catalog to make the final NIR catalog for the 4$\farcm$6$\times$4$\farcm$6 area centered on IRAS 22308+5812. A comparison between TIRSPEC and UKIDSS
 photometry was performed by plotting the differences of UKIDSS and TIRSPEC magnitudes with respect to TIRSPEC magnitudes (shown in Figure \ref{fig0}). We obtained differences of 0.00$\pm$0.22, -0.07$\pm$0.19, and
 0.00$\pm$0.24 mag in $J$-, $H$- and \ks-bands, respectively. Note that there could be variable sources which are possibly causing scatter in the difference between magnitudes obtained at two epochs.
 \subsection{{\it Spitzer} Infrared Spectrum}
{\it Spitzer} Infrared Spectrograph (IRS) archival data of the central bright source in the Sh2-138 region were obtained using {\it Spitzer} Heritage Archive. The basic calibrated data (BCD) images (Program ID: 1417,
 AOR key: 13048064, PI: IRS team) were used to extract the spectrum in low resolution channel (ch0: 5-15 $\mu m$). The final spectrum is obtained using the {\sc spice} package. The details of the
 extraction and reduction steps of IRS spectra are available in the {\it Spitzer}-IRS handbook\footnote[2]{http://irsa.ipac.caltech.edu/data/SPITZER/docs/irs/irsinstrumenthandbook/} (version 5.0).
\subsection{WISE data}
We utilized the publicly available archival WISE\footnote[3]{Wide Field Infrared Survey Explorer, which is a joint project of the University of California and the JPL, Caltech, funded by the NASA} \citep{wright10}
 images at 3.4 (W1), 4.6 (W2), 12 (W3), and 22 (W4) $\mu m$ as well as the photometric catalog. The resolution is 6$\arcsec$ for the first three bands and is 12$\arcsec$ for the 22 $\mu m$-band. The AllWISE point source
 catalog for 4$\farcm$6$\times$4$\farcm$6 area was obtained. 
\subsection{{\it Herschel} continuum maps}
{\it Herschel} \citep{pilbratt10} continuum maps were obtained at 70, 160, 250, 350, and 500 $\mu m$ using $Herschel$ Science Archive (HSA). The resolutions of these bands are 5$\farcs$8, 12$\arcsec$, 18$\arcsec$,
 25$\arcsec$, and 36$\arcsec$, respectively. We selected the processed level2$_{-}$5 MADmap images for the Photoconductor Array Camera and Spectrometer (PACS) 70 and 160 $\mu m$ bands, and the Spectral and Photometric
 Imaging Receiver (SPIRE) extended images at 250, 350, and 500 $\mu m$ bands, observed as part of the proposal name: OT2\_smolinar\_7. The 70--160 $\mu m$ maps are calibrated in the units of Jy pixel$^{-1}$, while
 the images at 250--500 $\mu m$ are in the surface brightness unit of MJy sr$^{-1}$. The plate scales of the 70, 160, 250, 350, and 500 $\mu m$ images are 3.2, 6.4, 6, 10, and 14 arcsec pixel$^{-1}$, respectively.
 We obtained 15$\arcmin\times$15$\arcmin$ maps at 70--500 $\mu m$ centred on $IRAS$ source. However, the analysis of temperature and column density maps is presented only for 9$\arcmin\times$9$\arcmin$ area.
\subsection{SCUBA 850 $\mu m$ continuum map}
Sub-mm Common-User Bolometer Array \citep[SCUBA;][]{holland99} 850 $\mu m$ continuum map (beam $\sim$14$\arcsec$) of the Sh2-138 region was obtained from the SCUBA legacy survey \citep{francesco08}. The observations
 were performed using the James Clerk Maxwell Telescope (JCMT). One can find more details about 850 $\mu m$ map in the work of \citet{francesco08}.
\subsection{Radio continuum data}
Radio continuum map at 1.4 GHz (beam size $\sim$45$\arcsec$) was obtained from the NRAO VLA Sky Survey (NVSS) \citep{condon98} archive. We also utilized the Very Large Array (VLA) archival images at 4890 MHz
(Project: AR0346) and 8460 MHz (Project: AR0304).
\subsection{JCMT HARP $^{13}$CO ($J=3-2$) data}
$^{13}$CO ($J=3-2$) (330.588 GHz; beam $\sim$14$\arcsec$) data were retrieved from the JCMT archive (ID: M08BU18; PI: Stuart Lumsden). The observations were obtained in position-switched raster-scan mode of the Heterodyne
 Array Receiver Program \citep[HARP;][]{buckle09}. We utilized the processed integrated CO intensity map of the Sh2-138 region.
\subsection{Bolocam 1.1 mm image}
We have obtained Bolocam 1.1 mm continuum map of the Sh2-138 region. The effective FWHM Gaussian beam size of the image is 33$\arcsec$ \citep{aguirre11}.
\section{Morphology of the region}
\subsection{A multi-wavelength view of the Sh2-138 region}
\label{subsecmorp}
\begin{figure*}
\includegraphics[width=\textwidth]{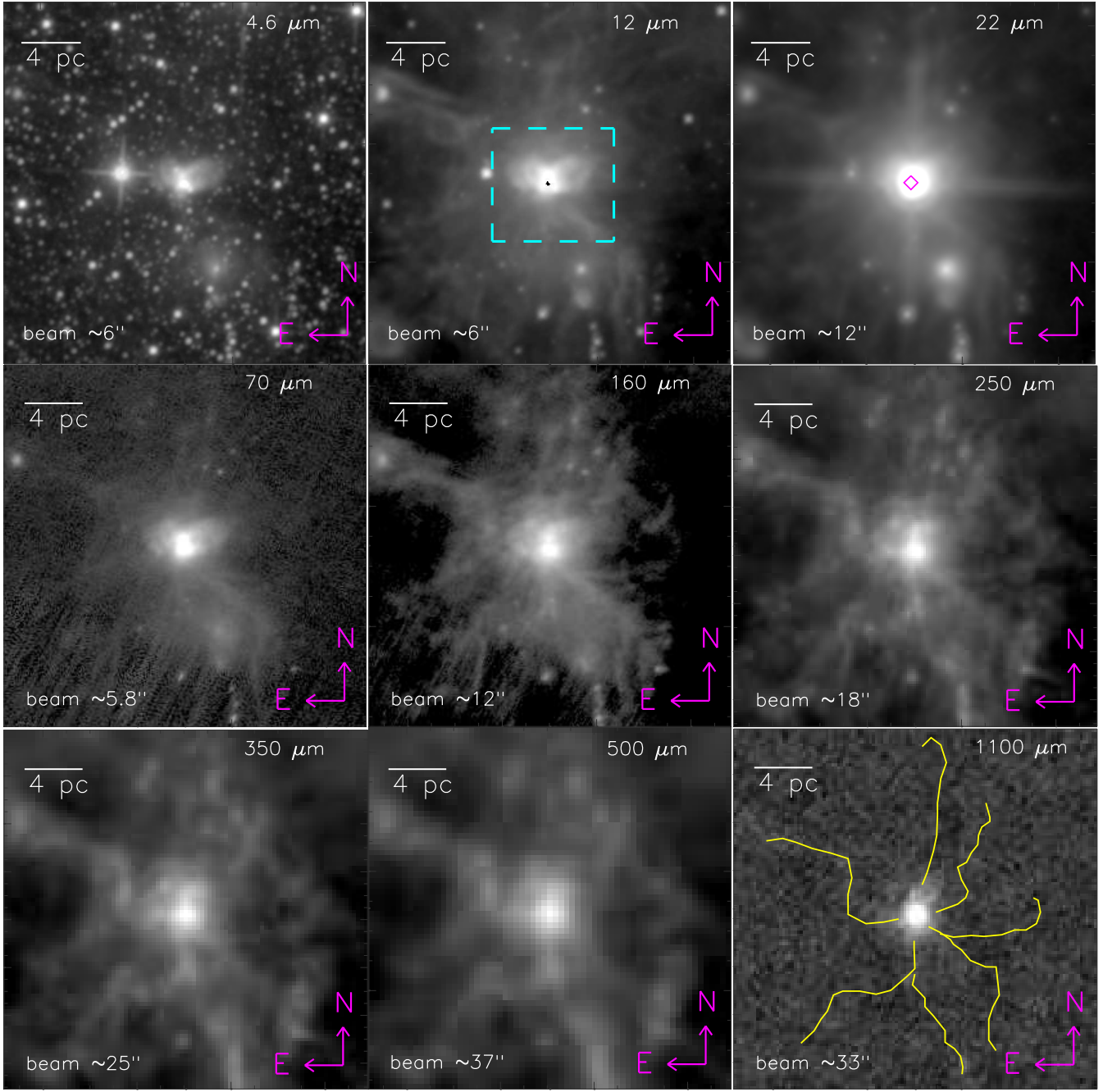}
\caption{A multi-wavelength view (at large scale 15$\arcmin$ $\times$ 15$\arcmin$; central coordinates: $\alpha_{2000}$ = 22$^{h}$32$^{m}$45$^{s}$.26, $\delta_{2000}$ = $+$58$\degr$28$\arcmin$20$\arcsec$.97)
 of the Sh2-138 region. The panels show images (at 4.6--1100 $\mu m$) from WISE (4.6--22 $\mu m$), {\it Herschel} Hi-GAL (70--500 $\mu m$), and BOLOCAM (1100 $\mu m$) (from left to right in increasing order).
 The scale bar on the top left shows a size of 4 pc at a distance of 5.7 kpc. The dashed cyan box in the panel of 12 $\mu m$ image is shown as a zoomed-in view in Figure~\ref{fig2}. The position of IRAS 22308+5812
 is marked by a diamond symbol ($\Diamond$) in 22 $\mu m$ image. The 12 $\mu m$ image is saturated near the $IRAS$ position. Filaments seen at 250 $\mu m$ map are shown by yellow curves in the panel of the 1100 $\mu m$
 map (see the text for more details).}
\label{fig1}
\end{figure*}

The spatial distribution of warm and cold dust emission in the Sh2-138 region at large scale (size~$\sim$ $15\farcm0  \times 15\farcm0$) is shown in Figure~\ref{fig1}. The features in the 4.6--70 $\mu m$ range trace
 the warm dust emission, while cold dust emission is traced in the 160--1100 $\mu m$ continuum images. Figure~\ref{fig1} illustrates the spatial morphology of the region, and the filamentary features are highlighted
 based on a visual inspection of the 250 $\mu m$ map (see curves in the 1100 $\mu m$ image panel in Figure~\ref{fig1}; hereafter {\it Herschel} filaments). All these filaments are several parsecs in lengths and appear
 to be radially emanating from the position of $IRAS$ source, revealing a `hub-filament' morphology \citep[e.g.][]{myers09}.

\begin{figure*}
\includegraphics[width=\textwidth]{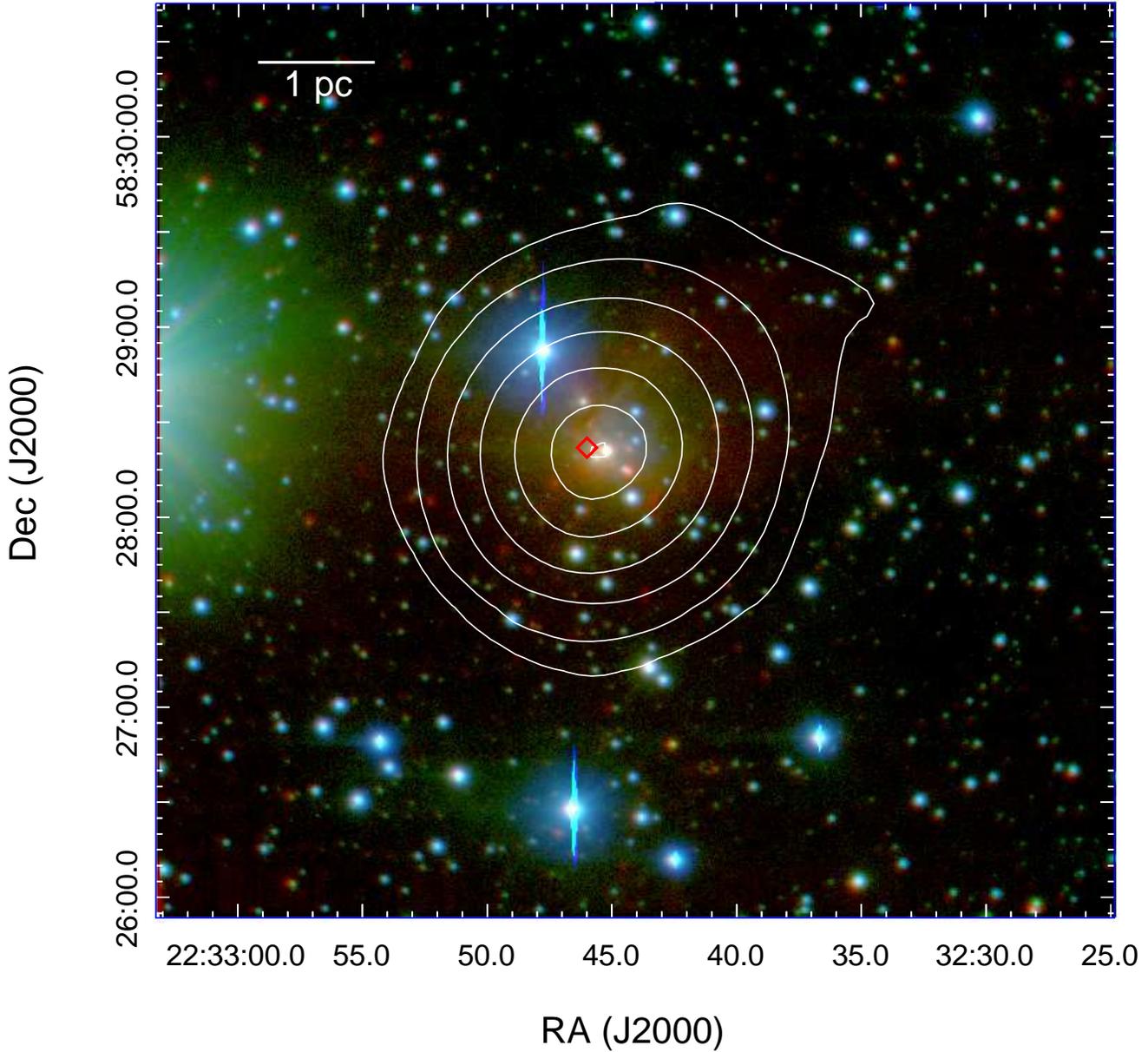}
\caption{A 3-colour composite image of the Sh2-138 region (size of the selected field 4$\farcm$6 $\times$ 4$\farcm$6; central coordinates: $\alpha_{2000}$ =22$^{h}$32$^{m}$44$^{s}$,
 $\delta_{2000}$ = $+$58$\degr$28$\arcmin$17$\arcsec$). The image is the result of the combination of TIRSPEC NIR \ks-band in red, $I$-band in green, and $V$-band in blue. The area of this image is shown by a
 dashed box in Figure~\ref{fig1}. Contours of NVSS 1.4 GHz radio emission in white colour are superimposed with 3$\sigma$, 20$\sigma$, 70$\sigma$, 200$\sigma$,  400$\sigma$, 650$\sigma$, and 850$\sigma$;
 where 1$\sigma\sim$1.56 mJy beam$^{-1}$. The position of IRAS 22308+5812 is marked by a diamond symbol ($\Diamond$). The scale bar at the top-left corner indicates 1 pc at a distance of 5.7 kpc.} 
\label{fig2}
\end{figure*}
\begin{figure*}
\includegraphics[width=0.9\textwidth]{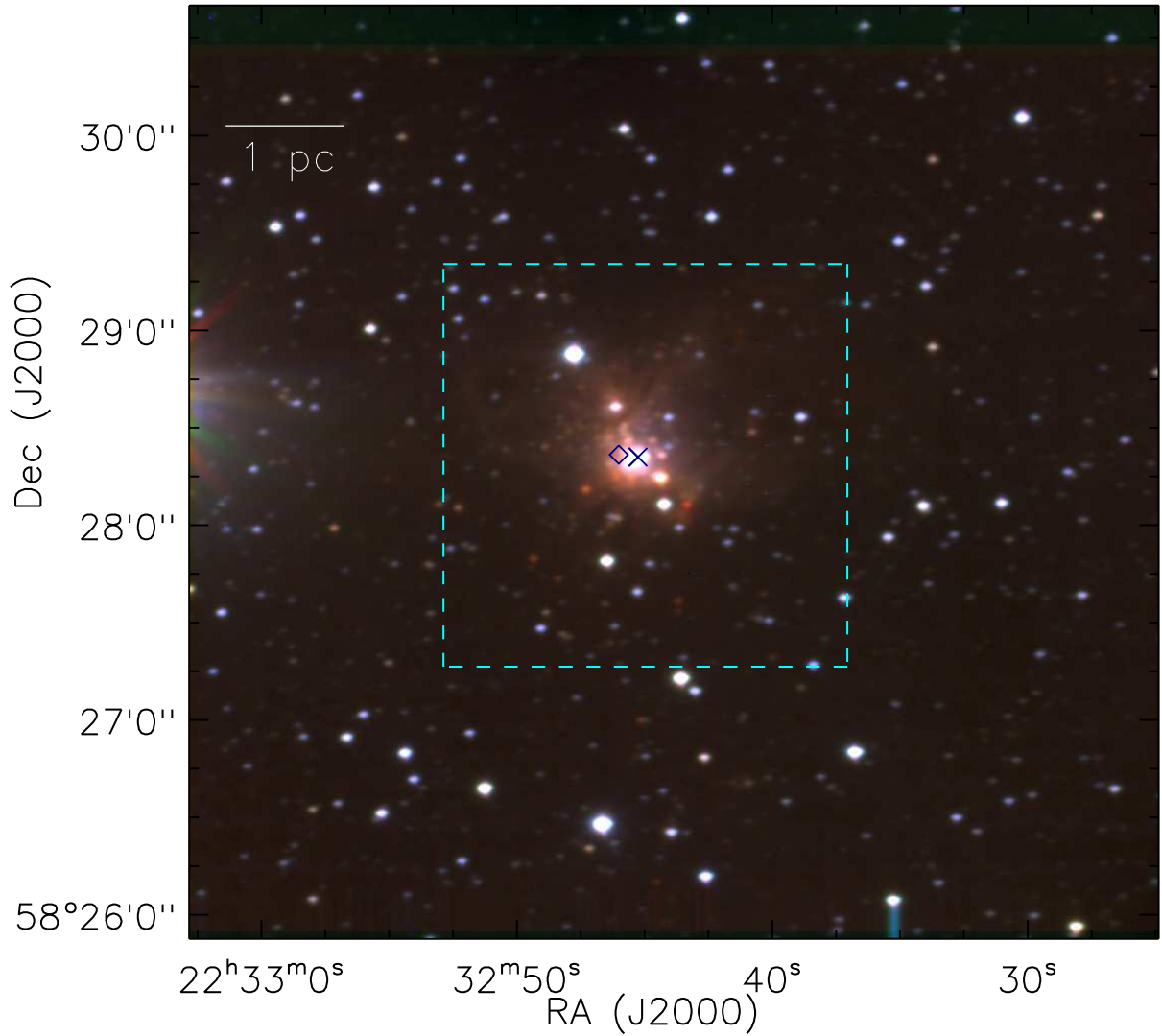}
\caption{TIRSPEC NIR colour composite images (\ks-band: red, $H$-band: green, and $J$-band: blue) in log scale. The area of this image is shown by a dashed box in Figure~\ref{fig1}.
 The position of IRAS 22308+5812 is marked by a diamond symbol ($\Diamond$). A bright source, located at the centre, is marked by a `$\times$' symbol. Optical, NIR, and MIR spectroscopic observations
 of this source are presented in Figures~\ref{fig10}--\ref{fig13}. One parsec scale bar is shown at the top-left corner, assuming distance to the region of 5.7 kpc. The dashed cyan box in the image is shown as
 a zoomed-in view in Figure~\ref{fig4}. }
\label{fig3}
\end{figure*}
\begin{figure*}
\includegraphics[width=0.8\textwidth]{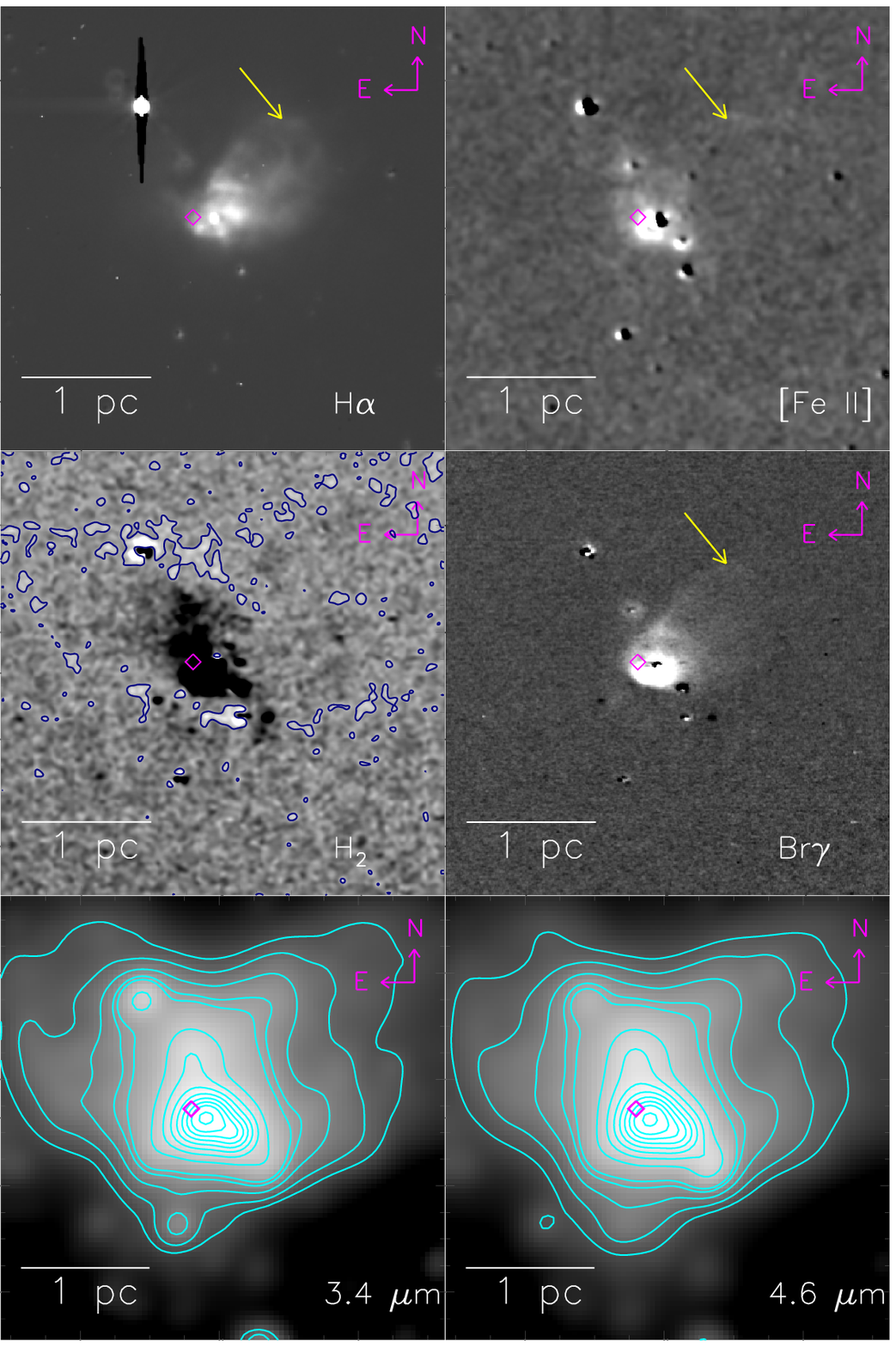}
\caption{Continuum-subtracted narrow-band and WISE 3.4 and 4.6 $\mu m$ images of the Sh2-138 region (size of the selected field $\sim 2$\farcm$07 \times 2$\farcm$07$; centered at $\alpha_{2000}$ = 22$^{h}$ 32$^{m}$ 45$^{s}$,
 $\delta_{2000}$ = +58$\degr$ 28$\arcmin$ 18$\farcs$4). The area of these images is shown by a dashed box in Figure~\ref{fig3}. Top left: $H\alpha$, top right: $[Fe$ {\sc ii}$]$, middle left: $H_2$, middle right: $Br\gamma$,
 bottom left: WISE 3.4 $\mu m$, and bottom right: WISE 4.6 $\mu m$. WISE 3.4 and 4.6 $\mu m$ contours are overplotted with the levels of 2, 3, 5, 8, 10, 20, 30, 40, 50, 60, 70, 80 and 95 \% of the peak values on the respective
 images. In all the panels, diamond symbol ($\Diamond$) marked the position of $IRAS$ source as shown in Figure~\ref{fig3}. The yellow arrows represent the detected faint features in the region. In order to enhance the faint
 features, the $H_2$ and $[Fe$ {\sc ii}$]$ maps are subjected to median filtering with a width of 6 pixels and smoothened by 6 pixels $\times$ 6 pixels. We have also drawn contours on $H_2$ image, above 1-$\sigma$ background
 level to bring out the faint features seen in the image. The scale bar indicates 1 pc at a distance of 5.7 kpc.}
\label{fig4}
\end{figure*}
The colour composite image made using $VI$\ks~bands ($V$: blue, $I$: green, and \ks: red) is shown in Figure~\ref{fig2}. The NVSS 1.4 GHz radio contours are also superimposed on the image to show the extent of the
 ionized region. One can infer from this image that the radio emissions and NIR \ks-band nebulosity are found around the bright source near the $IRAS$ source position. The diffuse nebulosity associated with the region is also
 seen in the NIR 3-colour image (see Figure~\ref{fig3}; $J$: blue, $H$: green, and \ks: red).

In Figure~\ref{fig4}, we present continuum-subtracted narrow-band images ($H\alpha$: 0.6563 $\mu m$, $[Fe$ {\sc ii}$]$: 1.644 $\mu m$, $H_2$: 2.122 $\mu m$, and  $Br\gamma$: 2.166 $\mu m$) and WISE 3.4 and 4.6
 $\mu m$ images of the Sh2-138 region. The diffuse emission features are clearly visible in the $H\alpha$ and $Br\gamma$ images, however the features present in the $[Fe$ {\sc ii}$]$ and $H_2$ maps are faint. We observed
 a spatial correlation of detected features in the $H\alpha$ and $Br\gamma$ images which are elongated in the northwest direction. The emissions detected in the $H\alpha$ and $Br\gamma$ maps trace the distribution of ionized
 emission in the region. NVSS radio contours (coarse beam) are also extended in the northwest direction. To enhance the faint features seen in the $H_2$ and $[Fe$ {\sc ii}$]$ maps, we have median filtered them with a
 width of 6 pixels and smoothened by 6 pixels $\times$ 6 pixels. The $H_2$ features are very faint in our image. However, we have shown the $H_2$ emission contours above the 1-$\sigma$ background level to bring
 out the $H_2$ distribution towards the $IRAS$ source. A bipolar cavity-like feature is traced in the $H_2$ map (see Figure~\ref{fig4}). One can also find the similar features near the $IRAS$ source in WISE 3.4
 and 4.6 $\mu m$ images. Earlier \citet{qin08} detected bipolar molecular $^{12}$CO ($J=2-1$) outflows towards the $IRAS$ source in this region. Hence, we suggest that these faint features seen in $H_2$ image are probably
 originated by shocks.

\begin{figure}
\includegraphics[width=0.45\textwidth]{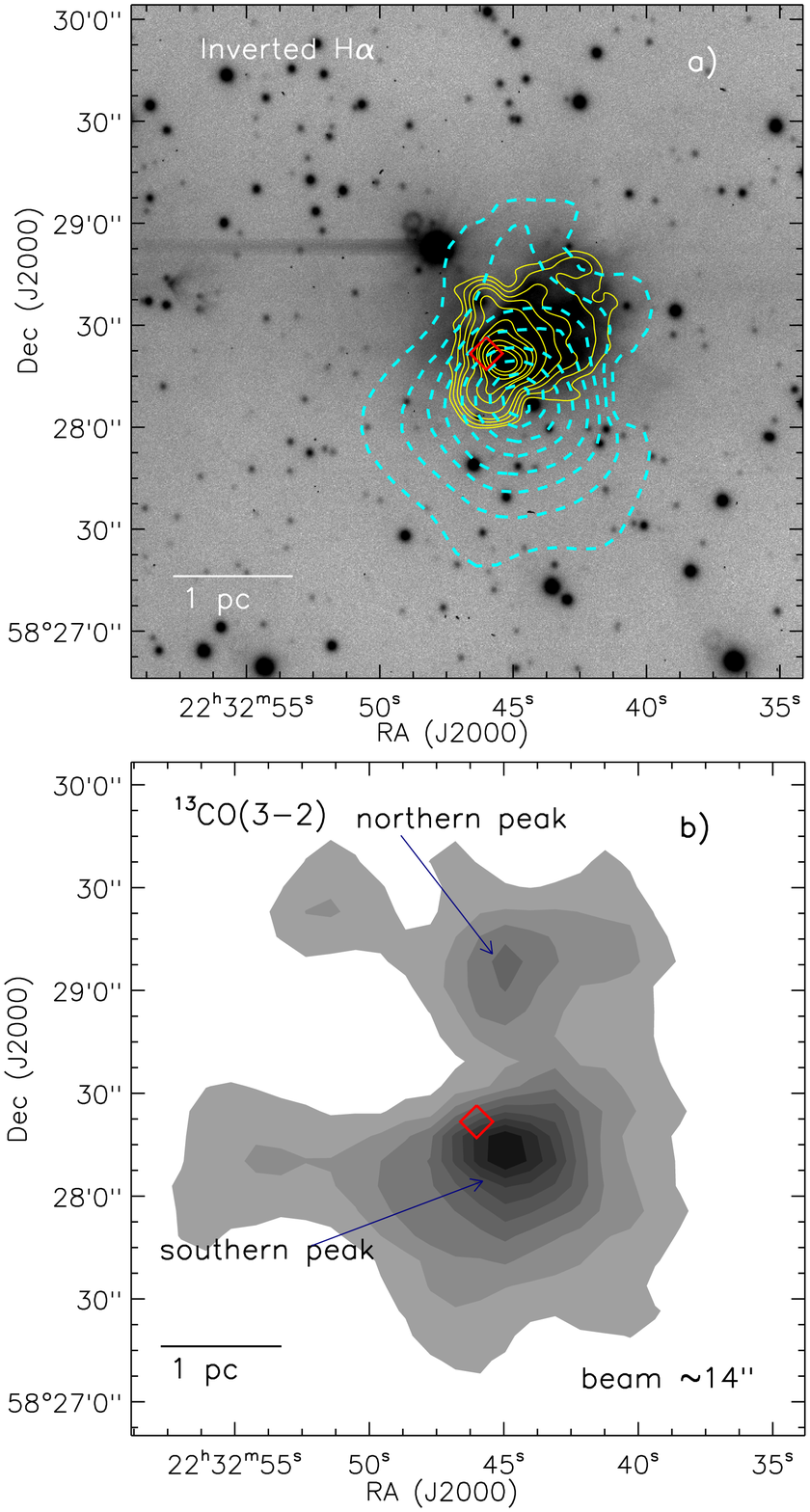}
\caption{The distribution of dust emission, molecular gas, and ionized emission in the Sh2-138 region. a) Overlay of GMRT 610 MHz radio contours (yellow lines) and SCUBA 850 $\mu m$ contours
 (dashed cyan lines) on the inverted grey-scale $H\alpha$ image (size of the selected field 3$\farcm$3 $\times$ 3$\farcm$3; central coordinates: $\alpha_{2000}$ =22$^{h}$32$^{m}$46$^{s}$.7,
 $\delta_{2000}$ = $+$58$\degr$28$\arcmin$25$\arcsec$). The 610 MHz radio contours are drawn at 12$\sigma$, 25$\sigma$, 50$\sigma$, 80$\sigma$, 150$\sigma$, 250$\sigma$, 300$\sigma$, 400$\sigma$,
 500$\sigma$, and 600$\sigma$, where 1$\sigma$ $\sim$74.0 $\mu$Jy beam$^{-1}$. SCUBA 850 $\mu m$ contours are also overlaid on the image with the levels of 2.572 Jy/beam $\times$ (0.1, 0.2, 0.3, 0.4,
 0.55, 0.7, 0.85, 0.95). b) The contour map of integrated JCMT-HARP $^{13}$CO ($J=3-2$) emission in the range of $-$58 to $-$45 km s$^{-1}$. The contour levels are 10, 20, 30, 40, 50, 60, 70, 80,
 and 90\% of the peak value (i.e. 62.2 km s$^{-1}$). In both the panels, the other marked symbol is similar to that shown in Figure~\ref{fig3}.}
\label{fig5}
\end{figure}

In Figure~\ref{fig5}, we present the distribution of ionized emission, molecular $^{13}$CO ($J=3-2$) gas, and cold dust emission in the Sh2-138 region. Figure~\ref{fig5}a shows GMRT 610 MHz radio contours (in yellow;
 beam $\sim$5$\farcs$5$\times$4$\farcs$8) and 850 $\mu m$ dust continuum contours (in cyan), overlaid on the $H\alpha$ image. The spatial distribution of ionized emission and $H\alpha$ emission is very well correlated.
 Figure displays that the peak position of 850 $\mu m$ emission is shifted ($\sim$11$\arcsec$) with respect to the peak position of 610 MHz radio emission. 

The extent of molecular cloud associated with the Sh2-138 region is revealed in Figure~\ref{fig5}b. In Figure~\ref{fig5}b, we find two peaks of molecular gas in the integrated CO intensity map (i.e. northern and southern)
 and the $IRAS$ source is located close to the southern peak of CO map. In summary, the southern CO emission region is associated with the dust and radio continuum emissions.
 
All together, we find that the CO and dust condensation associated with the $IRAS$ source is located at the junction of filaments, where the signature of active star formation (i.e. outflow) is evident.
\subsection{Radio morphology and physical parameters}
\label{sec_radio_morphology}
\begin{figure} 
\includegraphics[width=0.45\textwidth]{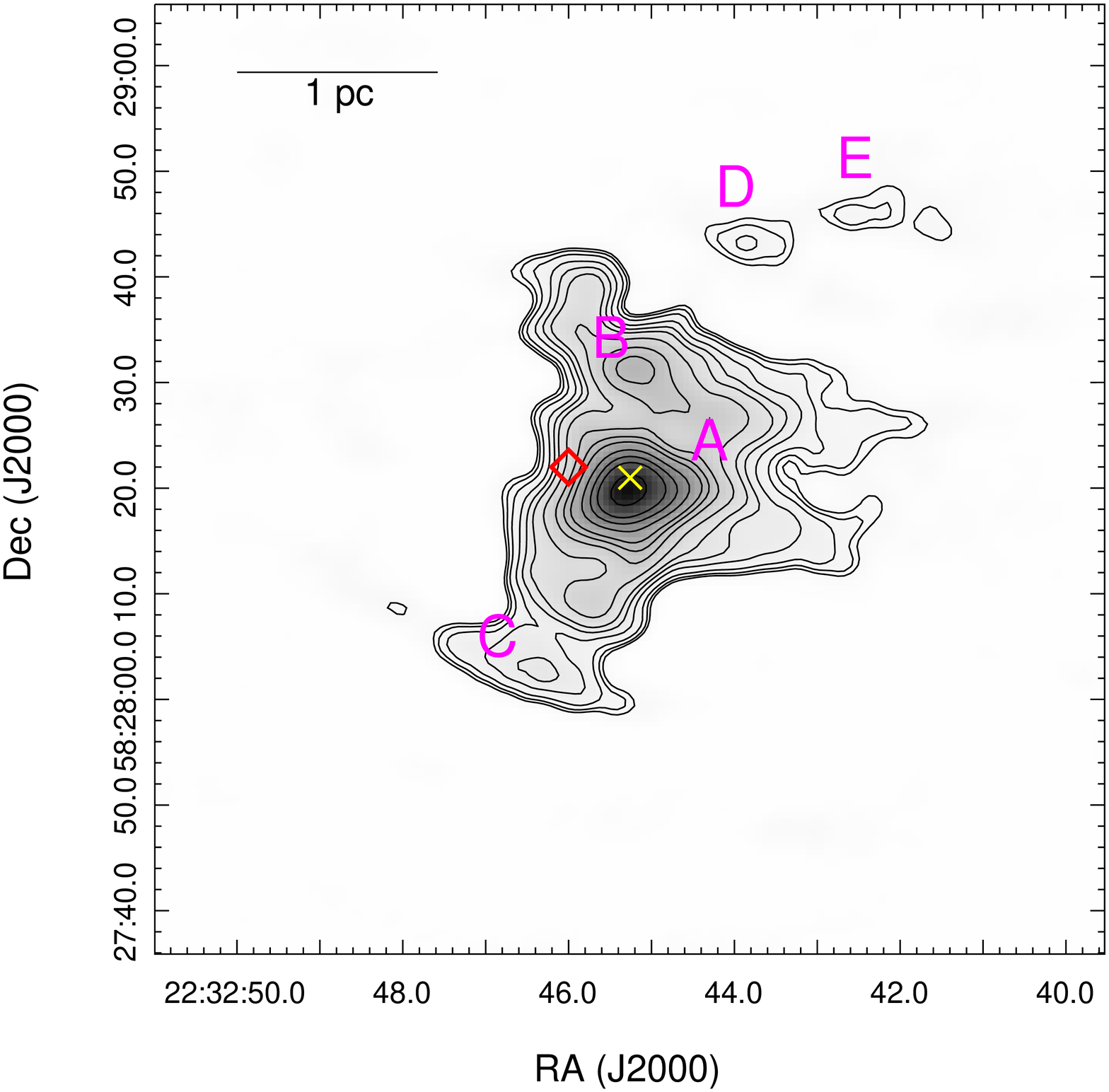}
\caption{The high resolution GMRT radio continuum contours map at 1280 MHz (beam $\sim$3$\farcs$4 $\times$ 2$\farcs$3). The contours are shown with levels of 12$\sigma$, 15$\sigma$, 20$\sigma$,
 30$\sigma$, 40$\sigma$, 50$\sigma$, 70$\sigma$, 100$\sigma$, 125$\sigma$, 150$\sigma$, 200$\sigma$, 250$\sigma$, 300$\sigma$, and 400$\sigma$, where 1$\sigma$ $\sim$71.7 $\mu$Jy/beam.
 The 1280 MHz map traces at least five peaks, which are marked by A, B, C, D, and E (see Table~\ref{table2}).}
\label{fig6}
\end{figure}

The presence of H {\sc ii} region is traced using NVSS 1.4 GHz emission (see Figure~\ref{fig2}), however NVSS radio map cannot provide more insight into the small-scale morphology of the Sh2-138 H\,{\sc ii} region
 due to coarse beam size. To examine the detailed compact features present in the H {\sc ii} region, we obtained high resolution GMRT radio maps at 610 MHz (beam $\sim$5$\farcs$5$\times$4$\farcs$8) and 1280 MHz
 (beam $\sim$3$\farcs$4$\times$2$\farcs$3 or 0.094 pc $\times$ 0.063 pc). The 610 MHz map was presented in Section~\ref{subsecmorp}. The radio continuum map at 1280 MHz is shown in Figure~\ref{fig6}, which reveals at
 least five compact peaks (also referred as radio clumps in this work) that are designated as A, B, C, D, and E.

We estimated the Lyman continuum flux (photons s$^{-1}$) for all these five clumps using the radio continuum flux, following the equation given in \citet{moran83}:
\begin{equation}
    S_{Lyc} = 8 \times 10^{43}\left(\frac{S_\nu}{mJy}\right)\left(\frac{T_e}{10^4 K}\right)^{-0.45}\left(\frac{D}{kpc}\right)^2 \left(\frac{\nu}{GHz}\right)^{0.1}
\label{lyman_flux}
\end{equation}
where $S_\nu$ is the integrated flux density, $T_e$ is the electron temperature, $D$ is the distance to the source, and $\nu$ is the frequency. In the calculation, we assumed that the region is homogeneous and spherically
 symmetric, and each clump is powered by a single zero age main-sequence (ZAMS) source. The flux density ($S_\nu$) and the size of each clump were determined using the AIPS task {\sc jmfit}. The electron temperature,
 $T_e$, for these H {\sc ii} clumps was typically assumed to be 10$^4$ K, except clump `A' \citep{stahler05}. For clump `A', $T_e$ was found to be 9250 K using the model of \citet{mezger67} which is discussed in detail in
 the next paragraph. We estimated the spectral type of
 the powering star associated with each radio clump by comparing its Lyman continuum flux with the theoretical values given in \citet{panagia73}. The resultant spectral type of the powering star associated with each clump
 is given in Table \ref{table2}. In Table \ref{table2}, we have also listed `clump name', clump peak position, observed frequency, integrated flux, size of the clump, and Lyman continuum flux. We find that four clumps
 (B, C, D, and E) are associated with sources of spectral type of B (earlier than B0.5V). The radio clump `A' is ionized by  an O9.5V type source, which is in agreement with previously estimated spectral type using 4.89 GHz
 flux by \citet{fich93}. A comparison of the GMRT 1280 MHz map, optical, and NIR images suggests that there are no optical and NIR counterparts found for four radio clumps (i.e. B, C, D, and E). However, the clump `A' is found
 to be associated with at least three sources at the centre.
 
\begin{figure}
\includegraphics[width=0.45\textwidth]{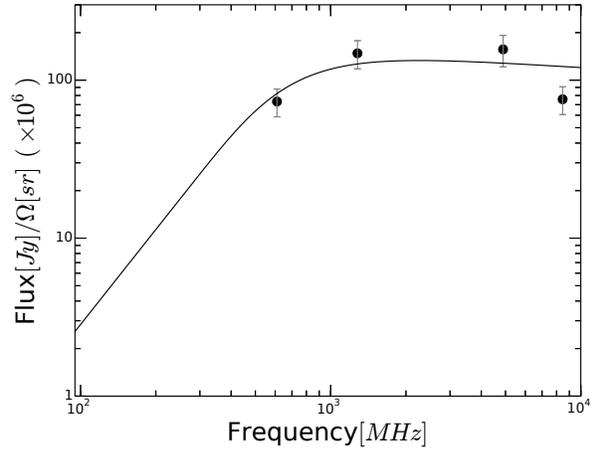}
\caption{Radio SED of the central radio clump (`A'). Black filled circles are the flux densities at 610, 1280, 4890, and 8460 MHz. The error bars are shown for a $\pm$10\% error limit. The black curve is the model fit
 to the flux values. The fluxes at 610 and 1280 MHz are obtained from GMRT and the remaining two points are taken from VLA.}
\label{fig7}
\end{figure}

Note that the central compact clump `A' is detected at four radio frequencies (i.e. GMRT 610 and 1280 MHz and VLA 4890 and 8460 MHz). Therefore, free-free emission spectral energy distribution (SED) fitting of the clump
 `A' is performed to estimate its physical parameters (such as emission measure, electron density, Str\"{o}mgren radius, and dynamical timescale). According to the model of \citet{mezger67}, the flux density due to
 free-free emission arising in a homogeneous and spherically symmetric region can be written as:
\begin{equation}
S_\nu = 3.07 \times 10^{-2} T_e \nu^{2}\Omega \left(1-e^{-\tau(\nu)}\right)
\label{radio_fit}
\end{equation}
\begin{equation}
\tau(\nu) = 1.643\times 10^5 a T_e^{-1.35}\nu^{-2.1}n_e^2l
\end{equation}
where, $S_\nu$ is the integrated flux density in Jansky (Jy), $T_e$ is the electron temperature of the ionized core in Kelvin, $\nu$ is the frequency in MHz, $n_e$ is the number of electrons in cm$^{-3}$, $l$ is the extent
 of the ionized region in parsec, $\tau$ is the optical depth, $a$ is the correction factor, and $\Omega$ is the solid angle subtended by the source in steradian. The parameter $n_e^2 l$ represents the emission measure
 (in cm$^{-6}$ pc), which is a measure of optical depth of the medium. In the calculations, we adopted the value of $a$ = 0.99 \citep{mezger67}. The observed radio fluxes of clump `A' were fitted by treating the
 temperature and emission measure as free parameters \citep[similar to ][]{omar02, vig14}. The model fit is sufficient to converge rapidly if observations are available at least one at optically thick regime (like 610 MHz) and another
 one at optically thin regime (like 4890 MHz). A chi-square minimization method was used in the fit to get the best estimate of the free parameters. The best fitted model returned a temperature of 9250$\pm$2000 K and an
 emission measure of 1.48$\pm$0.40$\times$10$^6$ cm$^{-6}$pc (see Figure~\ref{fig7}).

The knowledge of emission measure value and the linear extent ($l$) of the clump allows to estimate its electron density. The radio clump is comparatively more optically thin at 8460 MHz than other three observed radio frequencies.
 To obtain a better estimate of the extent of the clump, we calculated its linear extent ($l$) at 8460 MHz map, assuming a spherical morphology. The physical extent of the clump was found to be 0.29$\pm$0.05 pc for a  distance of
 5.7$\pm$1.0 kpc. Finally, the electron density was estimated to be 2250$\pm$400 cm$^{-3}$. Earlier, \citet{felli81} also obtained an electron density of 2500 cm$^{-3}$ for this H {\sc ii} region using the flux at
 4.99 GHz. Following the values tabulated in \citet{kurtz02}, our estimates of electron density ($\sim$2250 cm$^{-3}$) and of the linear extent of clump ($\sim$0.29 pc) correspond to a compact H {\sc ii} region.

\begin{table*}
\centering
\caption{Properties of the five radio clumps detected in GMRT 1280 MHz high resolution radio map.}
\begin{tabular}{@{}cccccccl@{}}
\hline
Clump& RA (J2000) &    Dec (J2000)   & Obs. freq. &    Integrated     &           Size                 &log (S$_{Lyc}$) &   Sp. type  \\
     &  (deg)     &     (deg)        &  (MHz)     &    flux (mJy)     &                                & (photons/sec)  &             \\
\hline
A   & 338.188500  &     58.472133    &   610.0    &  390.01$\pm$0.66  & 14$\farcs$7$\times$13$\farcs$6 &   48.000       &   O9.5      \\
    &             &	                 &  1280.0    &  276.74$\pm$0.63  &  9$\farcs$5$\times$7$\farcs$4  &   47.883       &   O9.5      \\
    &             &                  &  4890.0    &  310.53$\pm$0.73  &  9$\farcs$2$\times$8$\farcs$1  &   47.991       &   O9.5      \\
    &             &                  &  8460.0    &  369.31$\pm$0.50  & 14$\farcs$8$\times$12$\farcs$4 &   48.090       &   O9.0      \\
    &             &                  &            &                   &                                &                &             \\
B   & 338.188083  &	    58.475247    &  1280.0    &   97.33$\pm$0.69  & 10$\farcs$6$\times$7$\farcs$3  &   47.414       &   B0        \\
    &             &                  &  4890.0    &   46.08$\pm$0.83  & 11$\farcs$7$\times$7$\farcs$5  &   47.147       &   B0        \\
    &             &                  &            &                   &                                &                &             \\
C   & 338.193125  &	    58.467680    &  1280.0    &   39.46$\pm$0.78  & 13$\farcs$2$\times$7$\farcs$7  &   47.022       &   B0        \\
    &             &                  &            &                   &                                &                &             \\
D   & 338.182417  &	    58.478711    &  1280.0    &    9.94$\pm$0.50  & 10$\farcs$2$\times$5$\farcs$3  &   46.423       &   B0.5      \\
    &             &                  &            &                   &                                &                &             \\
E   & 338.176750  &     58.479464    &   610.0    &   11.65$\pm$0.42  & 14$\farcs$5$\times$8$\farcs$1  &   46.460       &   B0.5      \\
    &             &	                 &  1280.0    &    7.51$\pm$0.44  & 11$\farcs$3$\times$4$\farcs$1  &   46.301       &   B0.5      \\
\hline
\end{tabular}
\label{table2}
\end{table*}
The central massive source ionizes the surrounding medium and the ionization front expands until an equilibrium is established between the number of ionization and recombination. Theoretical radius of the H {\sc ii} region,
 called as Str\"omgren radius \citep{stromgren39}, assuming a uniform density and temperature, is given by:
\begin{equation}
R_S = \left(\frac{3S_{Lyc}}{4\pi n_0^2\beta_2}\right)^{1/3}
\end{equation}
where $R_S$ is the Str\"omgren radius, $n_0$ is the initial ambient density, and $\beta_2$ is the total recombination coefficient to the first excited state of hydrogen atom. The $\beta_2$ value for a temperature of 10$^4$ K
 is 2.60$\times$10$^{-13}$ cm$^3$ s$^{-1}$ \citep{stahler05}. 

\begin{figure*}
\includegraphics[width=\textwidth]{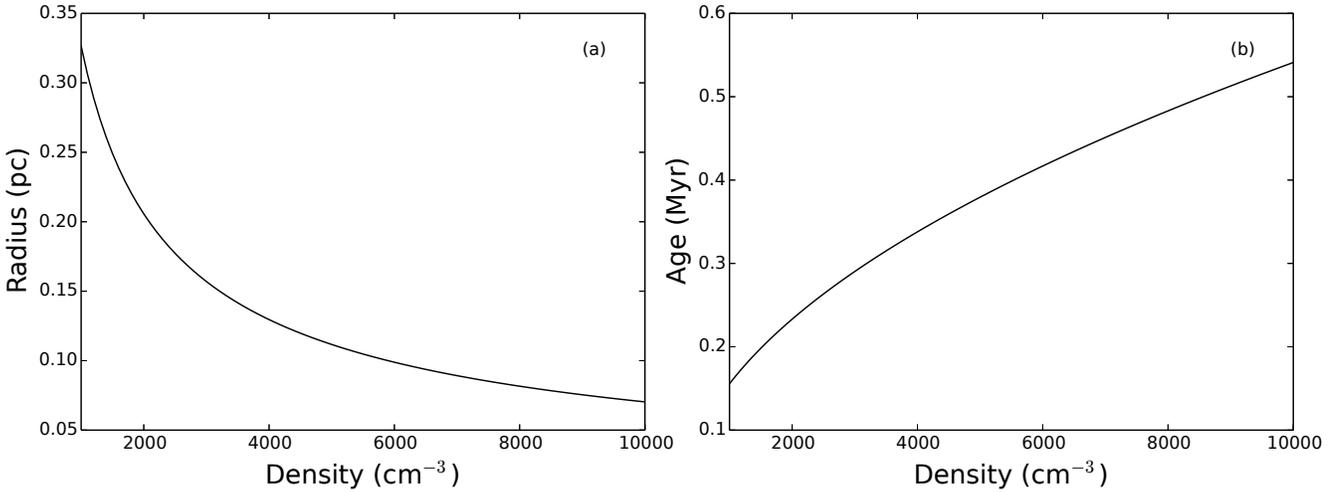}
\caption{a) Figure shows the variation of the Str\"{o}mgren radius with ambient density. b) Figure shows the variation of the dynamical age of the H {\sc ii} region with ambient density.}
\label{fig8}
\end{figure*}

During the second phase of the expansion of the H {\sc ii} region, a shock front is generated because of high temperature and pressure difference between the ionized gas and the surrounding cold material. This pressure gradient
 allows the shock front to propagate into the surroundings. When this phenomenon occurs, the radius of the ionized region at any given time can be written as \citep{spitzer78}:
\begin{equation}
R(t) = R_S \left(1 + \frac{7c_{II}t}{4R_S}\right)^{4/7}
\end{equation}
where c$_{II}$ is the speed of sound in H {\sc ii} region, which is assumed to be 11$\times$10$^5$ cm s$^{-1}$ \citep{stahler05}. The calculated age of the region is highly dependent on the initial value of the ambient
 density. Therefore, in Figure~\ref{fig8}, we plotted the variation of the Str\"{o}mgren radius and the age with ambient density, from 1000 to 10000 cm$^{-3}$ \citep[e.g. classical to ultra-compact H {\sc ii} regions;][]{kurtz02},
 for the central clump `A'. It can be seen in the figures that the Str\"{o}mgren radius and the age vary from 0.33 to 0.07 pc and from 0.16 to 0.54 Myr, respectively. It is to be noted that in the calculation of
 Str\"{o}mgren radius and dynamical age, the clump `A' is commonly assumed to be homogeneous and spherically symmetric which might not be always true. Also, the region is far enough to have unresolved companion(s) and hence,
 there might be contribution of Lyman continuum photons from those unresolved embedded massive stars in the cluster. Hence, the above calculated values (dynamical age and Str\"{o}mgren radius) can be considered as representative
 values for the clump `A'.
\subsection{Dust temperature and column density maps of the region}
In this section, we present the temperature and column density maps of the Sh2-138 region, generated using {\it Herschel} images. Following the same procedures as described in \citet{mallick15}, we generated the final temperature
 and column density maps by SED modeling of the thermal dust emission. In general, {\it Herschel} 70 $\mu m$ emission comes from UV-heated warm dust, and therefore the 70 $\mu m$ image is not used here. We adopted the following
 procedure to obtain the maps using {\it Herschel} 160--500 $\mu m$ fluxes. First, we transformed all the images into the same units (i.e. Jy pixel$^{-1}$), and then convolved the images to the resolution and pixel scale of the
 500 $\mu m$ image ($\sim$36$''$ resolution, 14$''$ pixel$^{-1}$) as it is the lowest among all the images. The background fluxes ($I_{bg}$), estimated from a relatively dark and smooth patch of the sky, were found to be 0.066,
 0.493, 0.297, and 0.127 Jy pixel$^{-1}$ for the 160, 250, 350, and 500 $\mu m$ images (area of the selected region $\sim$5$\farcm$8$\times$5$\farcm$8; central coordinates: $\alpha_{2000} \sim$ 22$^{\rm h}$31$^{\rm m}$17$^{\rm s}$,
 $\delta_{2000}\sim$ +58$^{\rm d}$32$^{\rm m}$09$^{\rm s}$), respectively. Finally, modified blackbody fitting was carried out on a pixel-by-pixel basis using the formula \citep{battersby11, sadavoy12, nielbock12, launhardt13}:
\begin{equation}
S_\nu (\nu) - I_{bg} (\nu) = B_\nu(\nu, T_d) \Omega (1 - e^{-\tau(\nu)})
\end{equation}
with optical depth formulated as:
\begin{equation}
\tau(\nu) = \mu_{H_2} m_{H} \kappa_\nu N(H_2)
\end{equation}
where $S_\nu (\nu)$ is the observed flux density, $I_{bg}$ is background, $B_\nu(\nu, T_d)$ is the modified Planck's function, $T_d$ is the dust temperature, $\Omega$ is the solid angle subtended by a pixel, $\mu_{H_2}$ is mean
 molecular weight, $m_H$ is the mass of hydrogen, $\kappa_\nu$ is the dust absorption coefficient, and $N(H_2)$ is the column density. In the calculations, we used $\Omega$ = 4.612$\times$10$^{-9}$ steradian
 (i.e. for 14$\arcsec\times$14$\arcsec$ area), $\mu_{H_2}$ = 2.8 and $\kappa_\nu$ = 0.1~$(\nu/1000~{\rm GHz})^{\beta}$ cm$^{2}$ g$^{-1}$, including a gas-to-dust ratio ($R_t$ =) of 100, with a dust spectral index of
 $\beta$\,=\,2 \citep[see][]{hildebrand83}.
\begin{figure}
\includegraphics[width=0.45\textwidth]{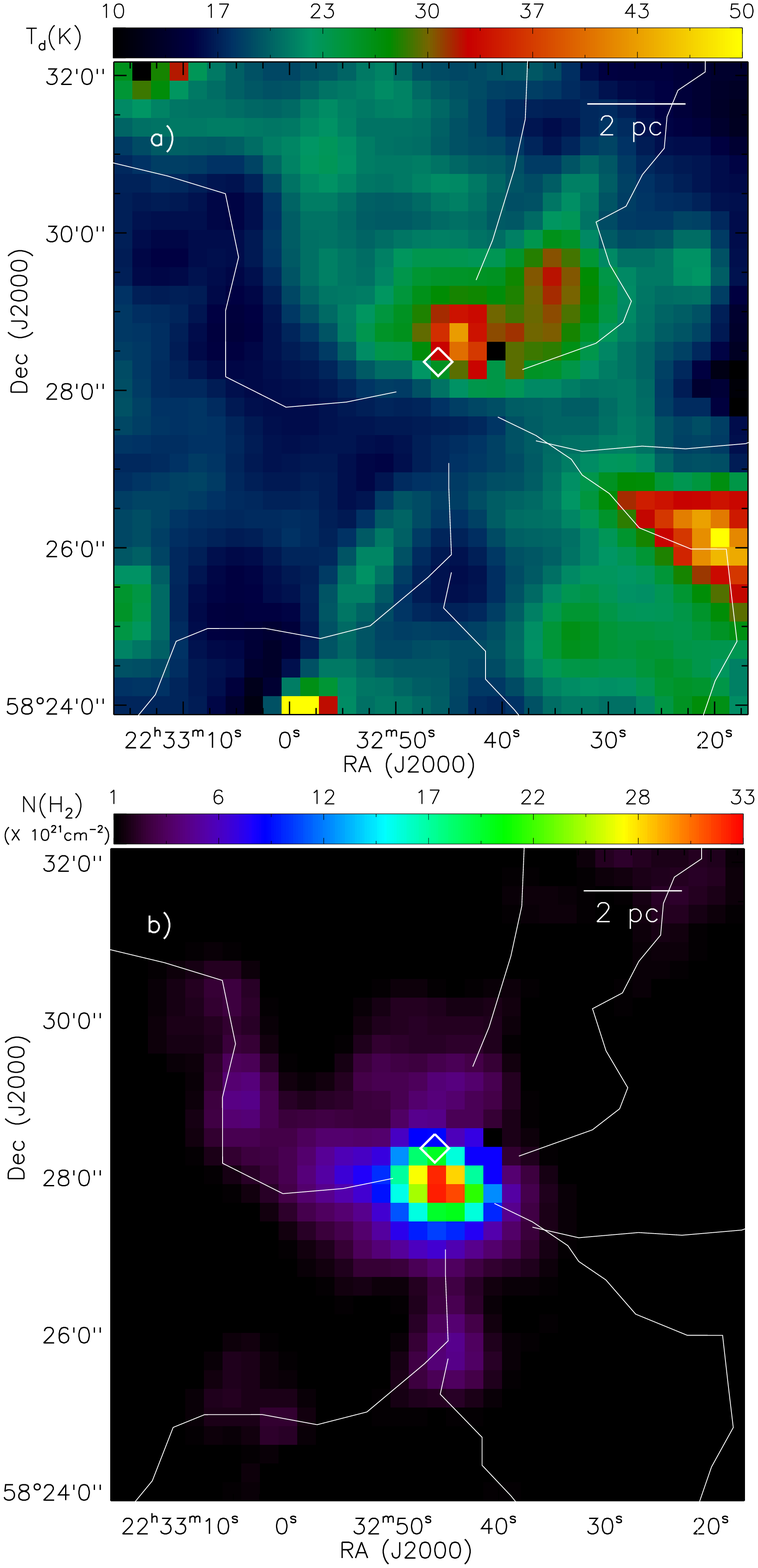}
\caption{a) {\it Herschel} dust temperature and b) column density ($N(\mathrm H_2)$) map of the Sh2-138 region (see text for details). The map also provides the information of extinction with
 $A_V=0.94 \times 10^{-21}~N(\mathrm H_2)$. The {\it Herschel} filaments (see Figure~\ref{fig1}) are also overplotted by solid curves in both the panels. The other marked symbol is similar to
 that shown in Figure~\ref{fig3}.}
\label{fig9}
\end{figure}
The resultant column density and temperature maps of the Sh2-138 region (angular resolution $\sim$36$\arcsec$; size~$\sim8\arcmin  \times 8\arcmin$) are shown in Figure~\ref{fig9}. {\it Herschel} filaments are also shown in
 Figure~\ref{fig9}. An isolated peak can be seen in the column density map (Figure~\ref{fig9}b), whose corresponding temperature from the temperature map is $\sim$30 $K$. To estimate the clump mass, we used the `clumpfind'
 software \citep{williams94} for measuring the total column density and the corresponding area of the clump seen in the column density map. The clump area was estimated to be $\sim$32 pc$^2$ ($\sim$218 pixels, where 1 pixel
 corresponds to $\sim$0.382$\times$0.382 pc$^2$), with a total column density of $\sim$1.16$\times$10$^{24}~cm^{-2}$. The mass of the clump can be estimated using the formula:
\begin{equation}
M_{clump} = \mu_{H_2} m_H Area_{pix} \Sigma N(H_2)
\end{equation}
where $\mu_{H_2}$ is assumed to be 2.8, $Area_{pix}$ is the area subtended by one pixel, and $\Sigma N(H_2)$ is the total column density estimated using `clumpfind'. Using these values, we obtained the total mass of the clump
 to be $\sim$3770 M$_\odot$. Such a large clump mass has been reported earlier in other massive star-forming regions, like M17 \citep[distance $\sim$1.6 kpc;][]{reid06} and NGC 7538 \citep[distance $\sim$ 2.8 kpc;][]{reid05}. 

In addition to the clump mass, the average visual extinction towards the region was also estimated from the column density value using the relation $\langle N(H_2)/ A_V \rangle = 0.94 \times 10^{21}$ molecules cm$^{-2}$ mag$^{-1}$
 \citep{bohlin78} and assuming that the gas is in molecular form. We calculated an average column density of $\sim$3.0$\times 10^{21}$ $cm^{-2}$ (in the central 5$' \times$5$'$ part of the column density map), which corresponds
 to visual extinction of $A_V$ $\sim$3.2 mag. This agrees well with the extinction value towards the region  found in literature \citep[$A_V$ $\sim$2.8 mag in][]{deharveng99}. 
\section{Spectral type of the central bright source}
As mentioned earlier, the radio clump `A' appears to be associated with at least three point sources including a bright optical and infrared source. In order to determine the spectral type of this central bright source, we
 present its optical, NIR, and $Spitzer$-IRS spectra in this section. In the slitless $H\alpha$ spectra (see Section~\ref{sec_ha_spec}), we identified two sources with enhanced $H\alpha$ emission including the central bright source.
\subsection{Optical spectrum}
\begin{figure*}
\includegraphics[width=\textwidth]{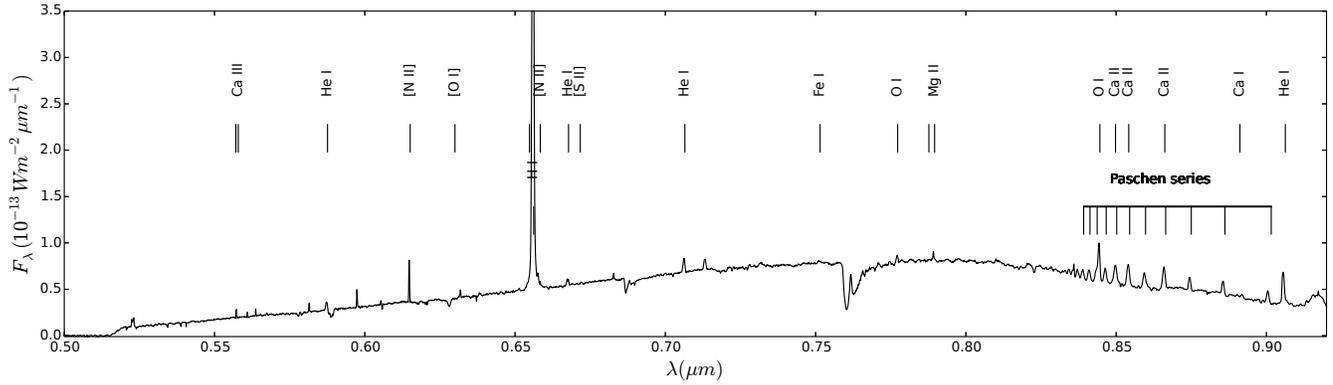}
\caption{Optical spectrum of the central bright central source (marked by a `$\times$' in Figure~\ref{fig3}). Apart from the hydrogen recombination lines, several ionized lines of Ca \& Mg and
 forbidden lines of [O {\sc i}], [N {\sc ii}] \& [S {\sc ii}] are also seen. The spectrum is similar to the spectrum of Herbig Be star, MWC 137 \citep{hamann92}.} 
\label{fig10}
\end{figure*}

The observed optical spectrum of the central bright source (5000-9200 {\AA}) is shown in Figure~\ref{fig10}. The spectrum is corrected for the nebular emission, which was obtained by averaging the spectra adjacent to the source
 spectrum, assuming that the density of the nebular emission region is uniform. Apart from the strong hydrogen recombination lines, emission lines of He {\sc i} at 5876 {\AA} \& 6678 {\AA}, [O {\sc i}] at 6300 {\AA} \& 6346 {\AA},
 [N {\sc ii}] doublet at 6548 {\AA} \& 6584 {\AA}, [S {\sc ii}] doublet at 6717 {\AA} \& 6731 {\AA}, and Ca {\sc ii} triplet at 8498 {\AA}, 8542 {\AA}, \& 8662 {\AA} are also seen in the spectrum. Presence of forbidden lines
 of [O {\sc i}] at 6300 \& 6364 {\AA}, and [S {\sc ii}] at 6717 \& 6731 {\AA} is generally seen in the spectra of Herbig Ae/Be stars and their low-mass counterparts (i.e. classical T Tauri stars (CTTS)). These forbidden lines are
 often used to infer  the presence of jets/outflows associated with the sources and originate only in low density conditions. Therefore, these lines are good tracers of excited low density material like jets/outflows
 \citep{finkenzeller85, corcoran98}.

The optical spectrum of the central bright source is similar to the spectrum of a Herbig Be star, MWC 137 \citep[see][]{hamann92}. These authors studied a total of 32 Herbig Ae/Be stars and found the Ca {\sc ii} triplet lines to
 be present in 27 sources, which is much higher than the detection rate in classical Be stars ($\sim$20\%). They also measured the equivalent widths of several characteristic lines of K, Fe, Mg, and Ca. In the case of MWC 137,
 the ratio of the equivalent widths of Ca {\sc ii} 8542 {\AA} to 8498 {\AA} lines was reported to be 1.16. It can be seen in our spectrum (Figure~\ref{fig10}) that the Ca {\sc ii} triplet lines are blended with the Paschen series
 lines. To get an estimate of the pure contributions of Ca {\sc ii} triplet lines, we first subtracted the contribution of the Paschen series lines' fluxes assuming a Gaussian profile, and then measured the equivalent width
 of the Ca {\sc ii} lines. All the hydrogen recombination lines are originated at the same part of the gas and hence, expected to have same velocity broadening. Gaussian profiles subtracted from blended lines were
 generated by keeping the velocity constant as it is obtained from isolated recombination lines. In this work, the
 ratio of equivalent widths of Ca {\sc ii} 8542 {\AA} to 8498 {\AA} lines is found to be 1.2$\pm$0.3. The equivalent width ratios of O {\sc i} 8446 {\AA} to 7773 {\AA} lines, and
 He {\sc i} 5876 {\AA} (blended) to 7065 {\AA} are 3.0$\pm$0.3 and 1.5$\pm$0.2, respectively. The ratio of the equivalent widths of these characteristic lines is dependent on the spectral type of the source, i.e, whether Herbig
 Ae star or Be star \citep{hamann92}. The analysis of optical spectra suggests that the central bright source is possibly a Herbig Be star.

We also estimated the electron density of the region from the ratio of [S {\sc ii}] 6716 {\AA} to 6731 {\AA} lines by comparing with the values given in \citet{canto80}. The [S {\sc ii}] 6716 {\AA} to 6731 {\AA} line ratio
 for our spectrum is 1.0$\pm$0.1, which corresponds to an electron density of $\sim$500$\pm$300 cm$^{-3}$ \citep[also see][]{osterbrock06}, assuming a temperature of 10000 K \citep[from ][]{stahler05}. Earlier, from
 their observed optical spectrum, \citet{deharveng99} had shown a variation of the electron density from 1000-200 cm$^{-3}$ from the immediate vicinity of the central bright source to the outer part of the nebula.
 \citet{martin02} also calculated the electron density of this region to be 768$^{+366}_{-207}$ cm$^{-3}$ using the [O {\sc iii}] 53 and 88 $\mu m$ lines' ratio from the ISO spectrum. The electron density of this region
 determined using the optical/infrared spectrum is always found to be lower than the electron density estimated from the radio analyses (2250$\pm$400 cm$^{-3}$ by us and $\sim$2500 cm$^{-3}$ by \citet{felli81}). However, the
 reason why the electron density value derived from optical/infrared spectrum is consistently lower than the value derived from radio analysis is not clear.
\subsection{{\it Spitzer}--IRS spectrum}
We present the $Spitzer$-IRS spectrum (5-15 $\mu m$) of the central bright source in Figure~\ref{fig11}. The emission lines of [Ne {\sc ii}] and [Ar {\sc ii--iii}] are seen in the spectrum, which suggest a
 presence of high energy photons in the circumstellar environment because the ionization of Ne and Ar, to the level of Ne {\sc ii} and Ar {\sc ii--iii}, requires high energy photons ($>$ 21 eV). These highly
 energetic UV/X-ray photons generally originate at the stellar chromosphere or accretion shocks from massive stars.

Apart from these ionized lines, several polycyclic aromatic hydrocarbon (PAH) emission lines are also seen in the spectrum. The relative strength of the PAH emission lines can be used to quantify the degree of
 ionization of the PAHs which further gives a clue about the incident radiation field and temperature. Generally, the ionized PAHs emit strongly at 6--9 $\mu m$ regime compared to the emission at 10--13
 $\mu m$ range \citep{allamandola99}.

\citet{sloan05} searched for any dependence of PAH emission features on their ionization fraction using the $Spitzer$-IRS spectra of four Herbig Ae stars, and found that the ratios of 7.7 and 11.3 $\mu m$ PAH
 features are correlated with the spectral types of sources in their sample. The ratio of 7.7 to 11.3 $\mu m$ PAH features was found to be 7.4 and 25.2 for spectral types of A0Ve and A5Ve, respectively, which
 implies a trend of decrease in ratio of PAH features for earlier spectral types of Herbig stars. For comparison, we have also measured fluxes at 7.7 $\mu m$ and 11.3 $\mu m$ PAH emission features from the
 $Spitzer$-IRS spectrum of our source and obtained fluxes of 43.2$\times$10$^{-15}$ Wm$^{-2}$ and 9.9$\times$10$^{-15}$ Wm$^{-2}$, respectively. It is difficult to determine the continuum level separately for
 7.7 and 8.3 $\mu m$ PAH features. \citet{sloan05} had therefore included 8.3 $\mu m$ PAH flux with 7.7 $\mu m$ flux in their study. Following \citet{sloan05}, we have also combined 8.3 $\mu m$ PAH flux with
 the flux at 7.7 $\mu m$ PAH feature to have a proper comparison. From the calculated flux, we obtained the ratio of fluxes at 7.7 and 11.3 $\mu m$ PAH features to be $\sim$4.8, which is lower than the values
 reported for Herbig Ae stars in \citet{sloan05}. It should be an indication that our source has an earlier spectral type than those four Herbig Ae stars reported in \citet{sloan05}, and hence, has a lower ratio
 of 7.7 and 11.3 $\mu m$ PAH features. We therefore suggest that the central bright source of the Sh2-138 region is possibly a Herbig Be star.
\subsection{NIR spectra}
\begin{figure}
\includegraphics[width=0.45\textwidth]{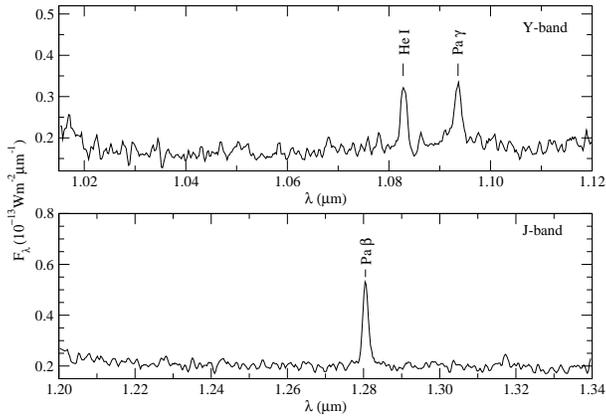}
\caption{TIRSPEC NIR $Y$-band (top panel) and $J$-band (bottom panel) spectra of the central bright source (marked in Figure~\ref{fig3}). Strong hydrogen lines and helium line at 1.083 $\mu m$ are
 seen in the spectra.\\\\\\\\}
\label{fig12}
\end{figure}

\begin{figure}
\includegraphics[width=0.45\textwidth]{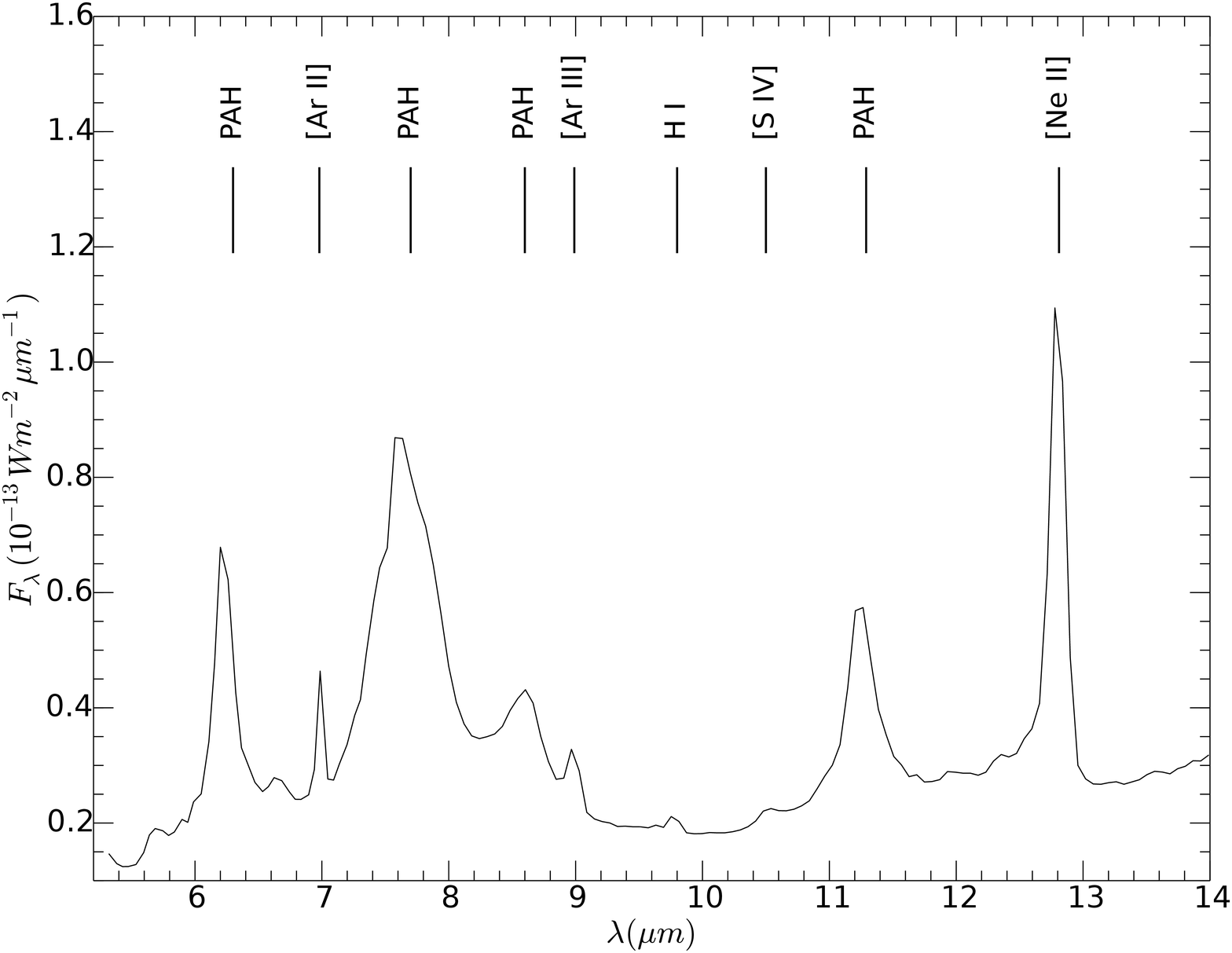}
\caption{Low resolution (Channel 0) {\it Spitzer}-IRS spectrum of the central bright source (marked in Figure~\ref{fig3}). Several PAH features and forbidden lines of
 Ar, Ne, and S lines are seen in the spectrum.}
\label{fig11}
\end{figure}

\begin{figure}
\includegraphics[width=0.45\textwidth]{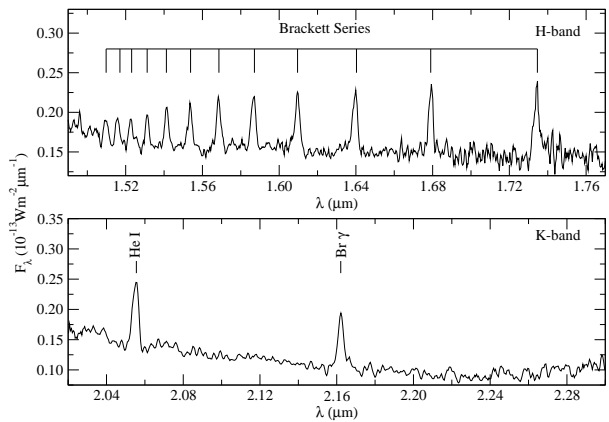}
\caption{TIRSPEC NIR $H$-band (top panel) and $K$-band (bottom panel) spectra of the central bright source (marked in Figure~\ref{fig3}). Strong Brackett lines are seen in both the spectra.
 Singlet counterpart of 1.083 $\mu m$ He {\sc i} line at 2.059 $\mu m$ is also seen in the $K$-band spectrum.}
\label{fig13}
\end{figure}

The NIR ($YJHK$-band) spectra of the central bright source are shown in Figures~\ref{fig12} and \ref{fig13}. Prominent hydrogen lines are mainly seen in all the NIR spectra. Apart from the hydrogen recombination lines, 
 strong He {\sc i} line at 1.083 $\mu m$ and its singlet counterpart at 2.058 $\mu m$ is also present. As mentioned in Section~\ref{subsecmorp}, faint $H_2$ features are detected around the $IRAS$ source. Therefore,
 the ro-vibrational line of $H_2$ (1-0) at 2.122 $\mu m$ is expected to be seen in the $K$-band spectrum. However, it is not detected possibly due to weak $H_2$ emission as seen in Figure~\ref{fig13}.

A qualitative estimation for the spectral type of the central bright source is carried out following the methods described in \citet{donehew11}. They performed UV, optical, and NIR spectroscopic study of 33 Herbig
 Ae/Be stars situated at different parts of the sky. They found a linear relationship between the accretion luminosity (L$_{acc}$) and the Br$\gamma$ luminosity (L$_{Br\gamma}$) for Herbig Ae stars in their sample.
 The $L_{Br\gamma}$ was found to be higher for Herbig Be stars compared to Herbig Ae stars for the same value of $L_{acc}$ \citep[see Figure 3 in][]{donehew11}.

\citet{donehew11} calculated the accretion luminosity of their sources using the formula:
\begin{equation}
L_{acc} = \frac{G M_\star \dot{M}}{R_\star}
\end{equation}
Where $M_\star$ is the stellar mass, $\dot{M}$ is the mass accretion rate, and $R_\star$ is the stellar radius. Following the same procedure, we estimated the L$_{\rm acc}$ of our source to be 5.3 L$_\odot$. In the
 calculation, we used the SED derived physical parameters of the source (stellar mass = 9.1 M$_\odot$, mass accretion rate = 7.8$\times$10$^{-7}$ M$_\odot$ yr$^{-1}$, and radius = 47.8 R$_\odot$; see Section~\ref{sec_yso_sed}
 for more details). We estimated the Br$\gamma$ flux of the source from the $K$-band spectrum to be 6.15$\pm$0.3$\times$10$^{-17}$ W cm$^{-2}$, which corresponds to a luminosity of 34$\pm$10 L$_\odot$. In case of the
 sources listed in \citet{donehew11}, we found that for an L$_{\rm acc}$ of $\sim$5.0 L$_\odot$, the ratio of L$_{\rm acc}$ to L$_{Br\gamma}$ is $\sim$5000 and $\sim$50 for CTTSs/Herbig Ae stars and Herbig Be stars, respectively.
 A further lower ratio of L$_{\rm acc}$ to L$_{Br\gamma}$ for our source ($\sim$0.16), implies that it should be at least a Herbig Be star.
 
 From the overall view on the optical, NIR, and MIR spectra, we conclude that the central bright source is most-likely a Herbig Be star.
\subsubsection{Extinction to the source}
Prominent $Pa\beta$ and $Br\gamma$ lines are detected in emission in the NIR spectra of the central bright source located near the $IRAS$ source (see Figures~\ref{fig12} and~\ref{fig13}). Hence, we used the observed
 ratio of these lines to estimate the extinction of the source. In order to infer the extinction value, we utilized the recombination theory and the observed flux ratio of $Pa\beta$ (1.282 $\mu m$) to $Br\gamma$ (2.166
 $\mu m$). The intrinsic flux ratio of $Pa\beta$ to $Br\gamma$ (i.e. (Pa$\beta$/Br$\gamma$)$_{int}$ = 5.89) was obtained for Case~B with T$_{e}$ = 10$^{4}$ K and n$_{e}$ = 10$^{4}$ cm$^{-3}$ from \citet{osterbrock06}.
 The differential reddening between $Br\gamma$ and $Pa\beta$ is given by, A$_{\gamma}$~$-$~A$_{\beta}$ = 2.5log$_{10}(Pa\beta/Br\gamma)_{obs}$ $-$ 2.5log$_{10}(Pa\beta/Br\gamma)_{int}$. The extinction at 1.282
 $\mu m$ can be estimated using, A$_{\beta}$~=~(A$_{\gamma}$~$-$~A$_{\beta}$)~[$(2.166~\mu m/1.282~\mu m)^{-1.9}$~-~1]$^{-1}$ \citep[e.g.][]{ho90}. We measured the ratio of $Pa\beta$ to $Br\gamma$ fluxes ($Pa\beta/Br\gamma$)$_{obs}$
 of 1.68 from our NIR spectra (see Figures~\ref{fig12} \& ~\ref{fig13}) and obtained A$_{\gamma}$ = 0.8 mag and A$_{\beta}$ = 2.16 mag using above relations. Subsequently, the visual extinction (A$_{V}$) of the source
 is estimated to be $\sim$7.0 mag using A$_{\gamma}$ and the extinction law of \citet{indebetouw05}. The deduced value of the extinction towards the central bright source is independent of distance. The difference between
 the foreground extinction (A$_V \sim$3.0 mag; see Section~\ref{sec_yso_selection}) and extinction towards this source indicates the presence of circumstellar material around the star.
\section{Young Stellar Population in the region}
\label{sec_yso_selection}
The study of young stellar sources and their distribution is important to understand the ongoing star formation in the region. In this section, we present the selection procedure of YSOs and analysis of their
 physical properties.
\subsection{Selection of YSOs}
We have identified YSOs in 4$\farcm$6$\times$4$\farcm$6 area of the Sh2-138 region using three different methods: (1) the NIR $J-H/H-K$ colour-colour diagram (CC-D), (2) the NIR $K/(H-K)$ colour-magnitude diagram (CM-D), and (3)
 the WISE CC-D. These methods are discussed in greater details in the following sections.
\begin{figure}
\includegraphics[width=0.47\textwidth]{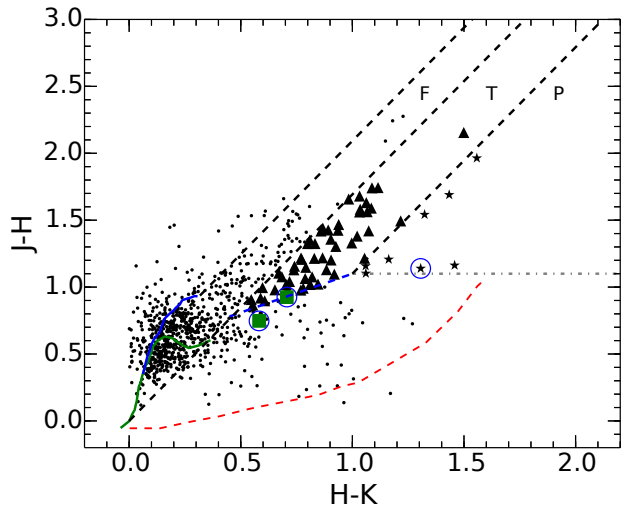}
\caption{NIR CC-D ($H-K~vs~J-H$) for the sources in the central 4$\farcm$6 $\times$ 4$\farcm$6 region of the Sh2-138. Solid triangles in the `T' region and the asterisks in the `P' region are Class II and Class I YSOs,
 respectively (see text). Two $H\alpha$ emission stars are marked by green squares. The solid curves represent the unreddened locus of MS dwarf stars (green) and giants (blue) \citep[from][]{bessell88}. The locus of
 the Herbig stars is shown in dashed red curve \citep{lada92}, while the locus of CTTS stars is shown by a dashed blue line \citep{meyer97}, and the dashed-dotted grey line drawn at $J-H$ = 1.1 from the tip of the CTTS locus
 is to decrease the contamination of Herbig Ae/Be stars in the `P' region. Three parallel black dashed lines are the reddening vectors drawn using the reddening laws from \citet{cohen81}. Three sources, identified in
 our \ks-band image (marked by blue circle) are arranged in a Trapezium-like configuration \citep{deharveng99}.}
\label{fig14}
\end{figure}
\subsubsection{NIR colour-colour diagram}
In Figure~\ref{fig14}, we show the NIR $J-H/H-K$ CC-D generated using the combined TIRSPEC and UKIDSS NIR catalog (see Section 3.1). The green curve in the CC-D represents the main-sequence (MS) locus, the solid blue curve is
 the locus of the giants \citep{bessell88}, the blue dashed line shows the locus of CTTS \citep{meyer97}, red dashed curve shows the locus of Herbig Ae/Be stars \citep{lada92} and the grey dashed-dotted line is drawn
 at $J-H$ = 1.1, extended from the tip of the CTTS locus. The parallel black dashed lines are the reddening vectors drawn from the base of the MS locus, turning point of the MS locus, and tip of the CTTS locus. All the magnitudes,
 colours, and loci of the MS, giants, CTTS and Herbig stars are converted to the Caltech Institute of Technology (CIT) system. The extinction laws A$_J$/A$_V$ = 0.265, A$_H$/A$_V$ = 0.155, A$_{K_{\rm s}}$/A$_V$ = 0.090 for
 the CIT system have been adopted from \citet{cohen81}.

The sources in the CC-D (Figure~\ref{fig14}) can be classified into three regions, namely, `F', `T', and `P' \citep[cf.][]{ojha04a, ojha04b}. The sources in the `F' region are generally considered as evolved field stars or Class~III
 YSOs (Weak-line T Tauri Stars). The sources in the `T' region are mainly Class~II YSOs \citep[CTTS;][]{lada92} with a large NIR excess and reddened early type MS stars with excess emission in the $K$-band \citep{mallick12}.
 The sources in the `P' region are Class~I YSOs with circumstellar envelopes. There may be an overlap of Herbig Ae/Be stars with the sources in the `T' and `P' regions, which generally occupy the place below the CTTS locus
 in the NIR CC-D \citep[for more detail see ][]{hernandez05}. Hence, to decrease any such contamination in the `P' region, we have considered only those sources which have $J-H~>~$1.1 mag. In this scheme, we have identified a
 total of 8 Class~I and 53 Class~II YSOs.
\begin{figure}
\includegraphics[width=0.47\textwidth]{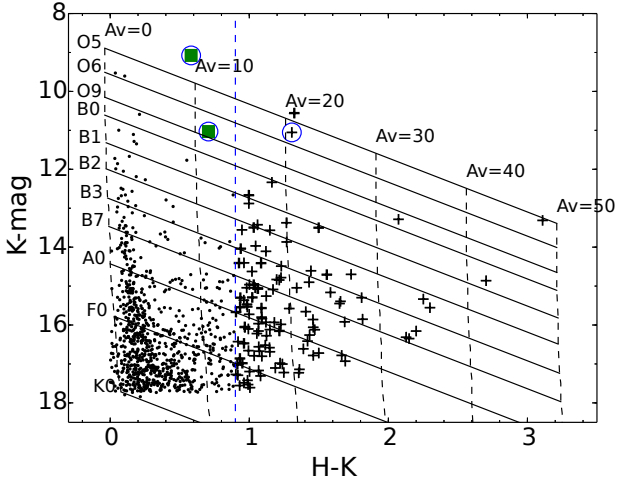}
\caption{NIR CM-D ($K/H-K$) of the sources detected in both $H$- and \ks-bands. Solid slanting parallel lines show the reddening vectors corresponding to the spectral type indicated in the left. Nearly vertical solid lines show
 the ZAMS loci for a distance of 5.7 kpc with foreground extinction of A$_{\rm v}$ = 0, 10,  20, 30, 40 and 50 mag, respectively. The sources with infrared excess, marked with `cross' symbols, are identified as
 candidate Class II/Class I YSOs. The cutoff value of $H-K~\sim~$0.9 is shown by blue dashed line. The remaining symbols/colours are similar to those in Figure~\ref{fig14}.}
\label{fig15}
\end{figure}
\subsubsection{NIR colour-magnitude diagram}
The NIR CM-D is a useful tool to identify a population of YSOs with infrared excess, which can be easily distinguishable from the MS stars. Figure~\ref{fig15} shows the NIR $K/H-K$ CM-D of the sources that are detected only
 in the $H$- and $K$-bands without any $J$-band counterparts. The CM-D of sources in the Sh2-138 region allows to identify additional YSOs. The black and red symbols in the CM-D are the same as used in the CC-D. Nearly vertical
 dashed black lines show the ZAMS loci for a distance of 5.7 kpc with foreground extinctions of A$_V$ = 0, 10, 20, 30, 40, and 50 mag. The slanted parallel lines represent the reddening vectors for different spectral types,
 drawn using the extinction laws of \citet{cohen81}.

It is to be noted that the UKIDSS catalog \citep[$H\sim$19.0; $K\sim$18.0;][]{lucas08} is deeper than the TIRSPEC catalog ($H\sim$18.0; $K\sim$17.8). Therefore, most of the faint and redder sources seen in the CM-D are observed
 only in the UKIDSS catalog. 

In Figure~\ref{fig15}, a low density gap of sources can be seen at $H-K\sim$~0.9 (marked by a blue dashed line). Therefore, red sources ($H-K>$~0.9) with infrared excess could be candidate Class II/Class I YSOs.
 The remaining sources with $H-K\leq$~0.9 are most-likely field stars. The colour criterion is consistent with the control field region (central coordinate: $\alpha_{2000} \sim$ 22$^{\rm h}$32$^{\rm m}$22$^{\rm s}$,
 $\delta_{2000}\sim$ +58$^{\rm d}$33$^{\rm m}$22$^{\rm s}$) where all the stars were found to have $H-K <$~0.9. Using the CM-D, a total of 114 sources have been detected as YSOs.
\subsubsection{MIR colour-colour diagram}
Additional YSOs are also identified using the WISE first three bands (W1, W2, and W3) magnitudes. The magnitudes with cc-flag of `D', `P', `H', and `O' (D: Diffraction spike; P: Persistence; H: Halo of nearby source; O: Optical
 ghost) for the first three bands were not considered in the analysis. We have removed the extragalactic sources, active galactic nuclei, shock objects, and PAH emission objects, following the colour criteria described in
 \citet{koenig08}. Finally, we identified two Class I type and one Class II type YSOs from the WISE (W1-W2)/(W2-W3) CC-D.

The selected YSOs using the above three methods might have overlap among themselves and hence, the YSOs detected in the NIR CC-D and CM-D were matched first. While matching, priority was given to the YSOs identified
in the NIR CC-D if the source is detected in both \citep[as it is done in][]{mallick13}, because the CC-D, constructed using 3 band magnitudes, provides more robust information about the Class of YSOs. No NIR counterparts of
 the WISE YSOs were found in the combined catalog of YSOs identified using the NIR CC-D and CM-D. Finally, we have identified a total of 149 YSOs in 4$'$.6$\times$4$'$.6 area (10 Class I, 54 Class II, and 85 infrared
 excess sources which could be candidate Class II/Class I YSOs), using the NIR CC-D or CM-D or MIR CC-D. The full catalog of YSOs is presented as supplementary material of this paper.
\begin{figure}
\includegraphics[width=0.45\textwidth]{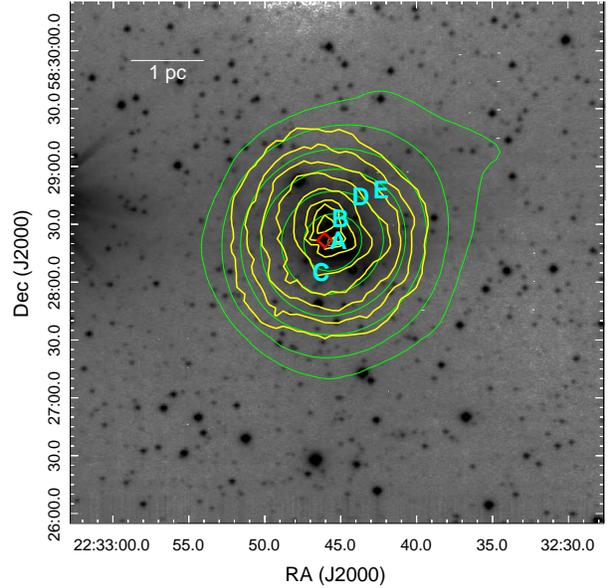}
\caption{YSOs surface density contours of the Sh2-138 region calculated for 20NN overlaid on the NIR \ks-band image. The surface density contours (in yellow) are drawn at 5, 7, 10, 15, 25, 40, 50, and 60 YSOs pc$^{-2}$. 
 We have also superimposed the NVSS 1.4 GHz radio contours (in green) at 3$\sigma$, 20$\sigma$, 70$\sigma$, 200$\sigma$, 400$\sigma$, 650$\sigma$, and 850$\sigma$; where $\sigma\sim$1.562 mJy/beam. The positions of the
 clumps detected in the 1280 MHz radio map are marked by `A', `B', `C', `D', and `E', and the other marked symbol is same as shown in Figure~\ref{fig3}.}
\label{fig16}
\end{figure}
\subsection{Surface density of YSOs}
\label{sec_nn_density}
The surface density analysis of identified YSOs in the Sh2-138 region is performed using the nearest neighbor (NN) method \citep[e.g.][]{casertano85, schmeja08, schmeja11}. We have performed 20NN surface density
 analysis because Monte Carlo simulations show that 20NN is adequate to detect cluster with 10 to 1500 YSOs \citep{schmeja08}. First, we divided the image (our selected 4$'$.6$\times$4$'$.6 FoV) into a $\sim$~3$\farcs$7
 (i.e. $\sim$~0.1 pc) regular grid. Subsequently, the distance to the 20$^{th}$ YSO at each point of the grid (i.e. $d_{20}$) was measured and then, the area of the circle was calculated by taking the measured distance as
 a radius. Finally, the surface density of YSOs per pc$^2$ ($\frac{20-1}{\pi d_{20}^2}$) was estimated at each point of the grid. Figure~\ref{fig16} shows the spatial correlation between YSO surface density and ionized
 emission. In the figure, the surface density contours and 1.4 GHz radio continuum emission are distributed nearly symmetric, and centered on the massive source (associated with clump `A'). Furthermore, the surface density
 map reveals an isolated cluster of YSOs in the Sh2-138 region, and all compact radio clumps seen at 1280 MHz are located within the cluster.
\begin{figure*}
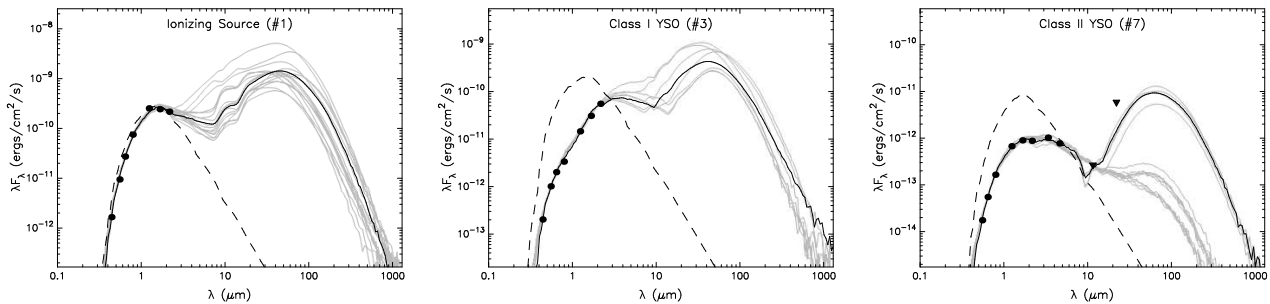
 
\begin{tabular}{ccc}
\includegraphics[width=0.3\textwidth]{ionizing.eps} &
\includegraphics[width=0.3\textwidth]{classI.eps} &
\includegraphics[width=0.3\textwidth]{classII.eps} \\
\end{tabular}
\caption{SED fitting for central brightest source (left), Class I (middle), and Class II (right) YSOs using the online tool of \citet{robitaille07} and their corresponding serial numbers in Table.\ref{table3} are also
 written in the bracket. The black dots are the data points, while the black triangles are the upper limit of fluxes. The solid black line is the best fit to the data points and the grey lines show the consecutive good
 fits for $\chi^2 - \chi^2_{min} <$ 3 (per data point). The dashed black line shows the best fit assuming that the observed fluxes are truly photospheric.}
\label{fig17}
\end{figure*}
\subsection{Spectral Energy Distribution}
\label{sec_yso_sed}
In this section, we present the results of SED modeling performed using an on-line SED modeling tool \citep{robitaille06,robitaille07} for a selected subset of YSOs that have photometric fluxes available in at least 5 bands. 
 The grids of YSO models, computed using the radiation transfer code of \citet{whitney03a, whitney03b}, are explained in \citet{robitaille06, robitaille07}. The models assume an accretion scenario with three different components 
 - (1) a pre-main sequence central star, (2) a surrounding flared accretion disk, and (3) a rotationally flattened envelope with cavities. The model grid has a total of 200,000 SED models and each model covers a range of stellar
 masses from 0.1 to 50 M$_{\odot}$. In the SED fitter tool, these models try to find out the best possible match for the given fluxes at different wavelengths, following the chi-square minimization, with distance and interstellar
 visual extinction (A$_{\rm V}$) as free parameters. The SED fitting tool requires at least three data points to model the observed SED, however the diversity of output model parameters can be constrained by providing more data
 points at longer wavelengths.
 
We selected 19 YSOs for the SED modeling that have detections in at least 5 bands. Additionally, 6 YSOs out of 19 have WISE fluxes. We treated these WISE fluxes as upper limits due to lower resolution. 
 We used A$_{\rm V}$ in the range of 3--40 mag and our adopted distance to the Sh2-138 region of 5.7$\pm$1.0 kpc, as input parameters for SED modeling. The A$_{\rm V}$ range is basically taken from the average foreground extinction
 to the extinction of the reddest source detected in our catalog. We only selected those models which satisfy the criterion: $\chi^{2}$ - $\chi^{2}_{best}$ $<$ 3, where $\chi^{2}$ is taken per data point. Note that the output
 parameters of SED modeling are not unique, hence, the computed parameters should be taken as representative values. Therefore, the weighted mean value of model fitted parameters for 19 selected YSOs are computed. Figure~\ref{fig17}
 shows the example model fits for the central massive source, a Class I YSO, and a Class II YSO, respectively. All 19 model fitted SEDs are presented as supplementary material of this paper. The weighted mean values of
 the stellar age, stellar mass, disk mass, disk accretion rate, envelope mass, stellar temperature, total luminosity, and extinction for 19 YSOs are listed in Table~\ref{table3}. Additionally, table also includes the positions of
 sources and $\chi^2_{min}$ values.

We noticed that the age of majority of YSOs lies between 0.1 and 4 Myr, with a mean age of $\sim$1 Myr (see Table \ref{table3}). Masses of YSOs show a range from 2 to 9 M$_\odot$ and majority of them (13 out of 19) have masses
 ranged between 2 and 6 M$_\odot$. Two sources have masses and ages $>$7 M$_\odot$ and $<$0.03 Myr, respectively, that are located in the central part of the cluster. The A$_V$ of YSOs varies between $\sim$3 to 9 mag. Two
 $H\alpha$ emission stars and a Class~I YSO show a high disk accretion rate of about 10$^{-6}$ to 10$^{-7}$ M$_\odot$ yr$^{-1}$. 
\subsection{Mass and Age spread from optical V/V-I colour-magnitude diagram}
\label{sec_VmI}
\begin{figure}
\centering
\includegraphics[width=0.45\textwidth]{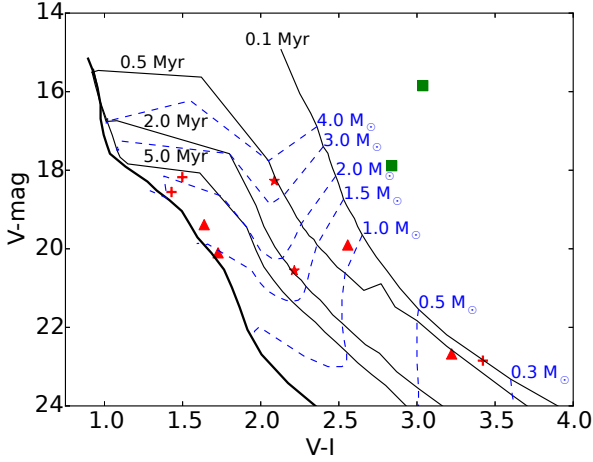}
\caption{YSOs detected in both $V$- and $I$-bands are plotted. The symbols are same as shown in Figures~\ref{fig14} and \ref{fig15} (asterisks: Class I YSOs; triangles: Class II YSOs; crosses: YSOs identified with
 infrared excess in NIR CM-D; green squares: sources with $H\alpha$ emission). Solid thick black line is the ZAMS locus, and thin black lines are PMS isochrones from 0.1--5 Myr adopted from \citet{siess00}. Blue
 dashed lines represent the evolutionary tracks for different masses. All the isochrones, ZAMS locus and evolutionary tracks are corrected for a distance of 5.7 kpc and foreground extinction of A$_{V}$ $\sim$3.0 mag.}
\label{fig18}
\end{figure}

The $V/V-I$ CM-D of YSOs, which are detected in optical $V$- and $I$-bands, is shown in Figure~\ref{fig18}. We have utilized optical photometry of YSOs to estimate their stellar mass and age. In general, the optical
 CM-D is a better tool to obtain stellar mass and age of YSOs than NIR CM-D, because optical fluxes of YSOs suffer less from their circumstellar emission than the fluxes in the NIR bands. In Figure~\ref{fig18},
 asterisks represent Class I YSOs, triangles represent Class II YSOs and crosses are sources with infrared excess identified in NIR CM-D. Though it is generally expected that Class I YSOs cannot be seen in optical
 bands, we have detected two of them possibly because they are located near the edge of `P' and `T' regions (see Figure \ref{fig14}) and with photometric uncertainty they can be either Class I or Class II YSOs. However, we
 left the nomenclature as Class I only, because technically they situated on the Class I side of the diagram. The green square symbols represent the enhanced $H\alpha$ emission sources which are detected in slitless
 $H\alpha$ spectra. In Figure~\ref{fig18}, we have overplotted the ZAMS locus for solar metallicity as well as the pre-main-sequence (PMS) isochrones for 0.1, 0.5, 2.0 and 5.0 Myr \citep[from][]{siess00}. The
 evolutionary tracks of PMS stars for 0.3, 0.5, 1.0, 1.5, 2.0, 3.0, and 4.0 M$_\odot$ are also shown in Figure~\ref{fig18}. All the isochrones, ZAMS locus and evolutionary tracks are corrected for a distance
 of 5.7 kpc and foreground extinction of A$_{\rm V}\sim$3.0 mag.

A wide spread of ages and masses can be noticed in Figure~\ref{fig18}. The sources with $H\alpha$ emission are found to be at age of $<$ 0.1 Myr, which are in agreement with the ages obtained from the SED modeling
 (see Table \ref{table3}). The masses of majority of YSOs are found in the range from 0.4 to 4.0 M$_\odot$, which also agree well with the  SED results. Note that we have corrected only the foreground extinction
 for isochrones, however the extinction due to local clouds and circumstellar material can also cause apparent spread in age and mass. Similar age and mass spreads were reported in other star-forming regions like
 Sh2-297 \citep{mallick12}, young open cluster Stock 8 \citep{jose08}, and NGC 1893 \citep{sharma07}. However, determining the ages of PMS stars is a rather difficult task and the age-spread obtained in CM-D can
 be explained due to variable extinction towards the individual sources, photometric variability due to presence of disk/accretion, spatially unresolved binaries, and scattered light from the disk with large
 inclination \citep{hillenbrand08, soderblom14}.
\subsection{The \ks-band Luminosity Function}
\label{sec_klf}
\begin{figure}
\centering
\includegraphics[width=0.45\textwidth]{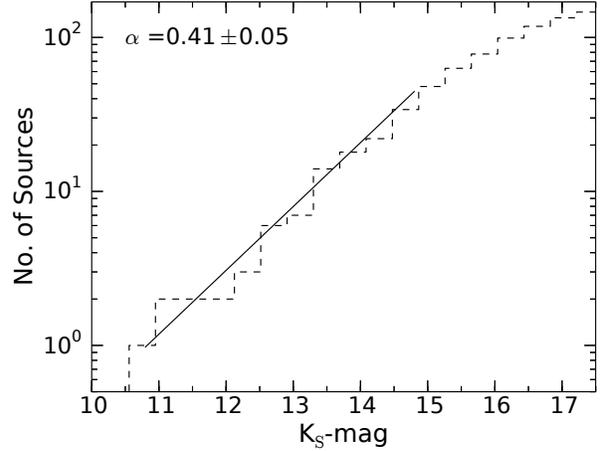}
\caption{The completeness-corrected cumulative KLF for the YSOs binned in 0.5 mag. The black straight line is the fit to the KLF in 11--14.5 mag range and the corresponding slope is given by $\alpha$.}
\label{fig19}
\end{figure}

An estimation of the age of a stellar cluster can be obtained from the \ks-band luminosity function \citep[KLF;][]{zinnecker93, lada95, vig14}. As pointed out by \citet{lada96}, the age of a cluster can be estimated
 by comparing its KLF to the observed KLFs of other young clusters. In the calculation, we assume that the luminosity function for a stellar cluster follows a power law. Therefore, one can define the KLF as
 $d$N(\ks)/$d$\ks $\propto$ 10$^{\alpha K_{\rm s}}$, where $\alpha$ is the slope of the power law. The KLF slope is estimated by fitting the cumulative number of YSOs in 0.5 \ks~magnitude bin, which is significantly higher
 than the errors associated with \ks-band magnitudes. We corrected the magnitude bins with corresponding completeness factor (see Section~\ref{sec_nir_phot}) prior to fitting of the slope. In Figure~\ref{fig19}, we show
 the completeness-corrected cumulative KLF of YSOs detected in the Sh2-138 region by dashed line, and the model fit to the KLF is shown by a solid line. The KLF is fitted for a magnitude range from 11.0 to 14.5 mag and the
 corresponding fitted slope is ($\alpha$ =) 0.41$\pm$0.05.

Similar values of $\alpha$ have been found by several authors in different high-mass star-forming regions. Recently, \citet{mallick14, mallick15} obtained $\alpha$ of 0.40$\pm$0.3 and 0.35$\pm$0.4 for NGC 7538
 (distance $\sim$2.65 kpc, age $\sim$1 Myr) and IRAS 16148-5011 (distance $\sim$3.6 kpc; age $\sim$1 Myr) star-forming regions, respectively. Earlier, \citet{lada95} and \citet{lada91} also found similar
 value of $\alpha$ for the Orion molecular cloud ($\sim$0.38, age $\sim$1 Myr). Comparable values of $\alpha$ are also found for several star-forming regions like, NGC 1893 cluster \citep[$\sim$ 0.34$\pm$0.07,
 distance $\sim$3.25 kpc; age $\sim$1-2 Myr;][]{sharma07}, and IRAS 06055+2039 cluster \citep[$\sim$0.43$\pm$0.09, distance $\sim$2.6 kpc; age $\sim$2-3 Myr;][]{tej06}. The age spread obtained for YSOs in the
 Sh2-138 region from the $V/V-I$ CM-D is similar to other massive star-forming region and the estimated KLF slope is in agreement with the slopes of several other young clusters. Noting the $\alpha$ values and ages
 of all these massive star-forming regions, we conclude that the mean age of the Sh2-138 cluster is possibly $\sim$1-2 Myr.
\subsection{The stellar mass spectrum}
\label{sec_JmH}
\begin{figure}
\centering
\includegraphics[width=0.45\textwidth]{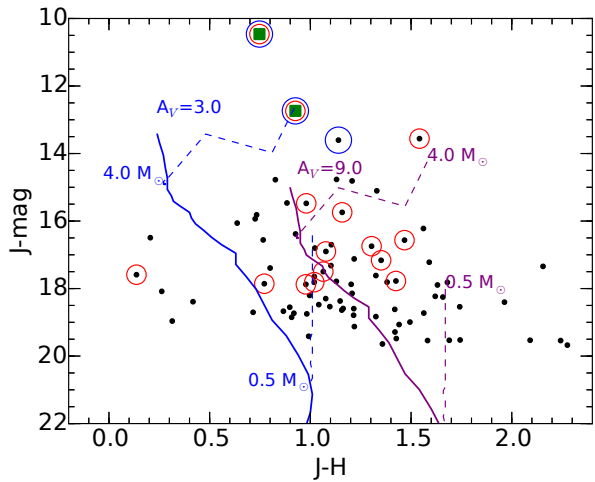}
\caption{$J/J-H$ CM-D for the YSOs detected both in $J$- and $H$-bands (black dots). The solid blue and maroon curves show ZAMS loci from \citet{siess00} for $A_V$ of 3.0 and 9.0 mag, respectively. The evolutionary tracks
 for 0.5 and 4.0 M$_\odot$ PMS stars are also shown with dashed curves for both the extinction values. The selected YSOs for which SED analysis was performed are shown with additional red circles. The remaining symbols are similar
 to those shown in Figure~\ref{fig14}.}
\label{fig20}
\end{figure}

In Section~\ref{sec_VmI}, we presented the mass and age spreads of YSOs using the $V/V-I$ CM-D. However, several YSOs identified in the Sh2-138 region do not have optical counterparts because of high extinction
 towards the region. Hence, we have constructed the NIR $J/J-H$ CM-D to obtain an estimate of the mass range for majority of YSOs. In the estimation of the stellar masses of YSOs using NIR magnitudes, only $J$- 
 and $H$-bands were considered. The NIR \ks-band magnitude is not used in this analysis because the circumstellar material of YSOs emits strongly in \ks-band regime compared to $J$- and $H$-bands and hence, inclusion
 of \ks-band flux in this analysis might lead to an overestimation of the stellar mass.
 The $J/J-H$ CM-D of YSOs detected in both $J$- and $H$-bands is shown in Figure~\ref{fig20}. We have overlaid ZAMS loci from \citet{siess00} for $A_V=$ 3.0 and 9.0 mag (see Section~\ref{sec_yso_sed} and
 Table~\ref{table3}) and the evolutionary tracks for 0.5 and 4.0 M$_\odot$ PMS stars \citep{siess00} for both these extinction values. All the loci and tracks are corrected for a distance of 5.7 kpc. The YSOs selected for the
 SED modeling are marked with red circles, and the enhanced $H\alpha$ emission sources are shown with green squares. Most of the YSOs are found to be well distributed in the mass range from 0.5--4.0 M$_\odot$, with
 $A_V$ ranged from 3.0--9.0 mag, which is consistent with the SED modeling results.
\begin{landscape}
\begin{table}
\centering
\caption{SED model output parameters of selected YSOs.}
\begin{tabular}{@{}cccccccccccc@{}}
\hline
Sr  & RA (J2000) & Dec (J2000) &   Log (Age)   &    Mass       &log (M$_{\rm disk}$)&log (\.M$_{\rm disk}$)& log (M$_{\rm env}$)& log (T$_\star$) & log (L$_{\rm tot}$) & A$_{\rm V}$ &$\chi^2$           \\    
No. & (deg)      &   (Deg)     &    (yr)       & (M$_\odot$)   &(M$_\odot$)         & (M$_\odot$yr$^{-1}$) &   (M$_\odot$)      &       (K)       &    (L$_\odot$)      &  (mag)      & (per data points) \\
\hline
\multicolumn{11}{c}{$H\alpha$ emission stars}\\
\hline
1  & 338.188570  & 58.472493   & 4.45$\pm$0.10 & 9.10$\pm$0.34 & -1.57$\pm$0.86 & -6.11 $\pm$0.86 &  1.31$\pm$0.54  & 3.73$\pm$0.03 & 3.23$\pm$0.08 & 4.36$\pm$0.38 & 3.27    \\  
2  & 338.190270  & 58.471976   & 4.94$\pm$0.34 & 5.62$\pm$1.06 & -1.72$\pm$0.88 & -7.16 $\pm$1.21 &  0.53$\pm$0.89  & 3.69$\pm$0.06 & 2.35$\pm$0.24 & 4.03$\pm$0.75 & 2.93    \\  
\hline
\multicolumn{11}{c}{Class I YSOs}\\
\hline
3  & 338.185070  & 58.470722   & 4.41$\pm$0.01 & 8.13$\pm$0.01 & -0.56$\pm$0.01 & -5.25$\pm$0.01  &  0.61$\pm$0.01  & 3.67$\pm$0.01 & 2.89$\pm$0.01 & 3.00$\pm$0.01 & 2.84    \\  
4  & 338.193315  & 58.471851   & 5.07$\pm$0.01 & 6.80$\pm$0.01 & -1.24$\pm$0.01 & -6.61$\pm$0.01  &  0.50$\pm$0.01  & 3.70$\pm$0.01 & 2.71$\pm$0.01 & 4.66$\pm$0.01 & 3.43    \\  
\hline
\multicolumn{11}{c}{Class II YSOs}\\
\hline
5  & 338.140670  & 58.497884   & 6.51$\pm$0.01 & 6.41$\pm$0.01 & -6.60$\pm$0.01 & -13.02$\pm$0.01 & -5.97$\pm$0.01  & 4.29$\pm$0.01 & 3.07$\pm$0.01 & 6.92$\pm$0.05 & 3.47    \\  
6  & 338.170170  & 58.474590   & 6.33$\pm$0.74 & 4.38$\pm$1.66 & -5.27$\pm$2.10 & -11.03$\pm$2.26 & -3.97$\pm$2.99  & 4.02$\pm$0.19 & 2.26$\pm$0.46 & 8.44$\pm$1.14 & 0.09    \\  
7  & 338.170830  & 58.444609   & 5.07$\pm$0.01 & 2.08$\pm$0.01 & -0.97$\pm$0.01 & -6.64 $\pm$0.01 &  0.03$\pm$0.01  & 3.64$\pm$0.01 & 1.58$\pm$0.01 & 4.72$\pm$0.01 & 4.66    \\  
8  & 338.171130  & 58.470674   & 5.46$\pm$1.06 & 4.06$\pm$2.68 &       --       &       --        & -1.80$\pm$3.78  & 3.71$\pm$0.11 & 1.74$\pm$0.87 & 4.47$\pm$1.09 & 0.77    \\  
9  & 338.173390  & 58.444698   & 5.59$\pm$1.16 & 4.15$\pm$2.65 &       --       &       --        & -2.21$\pm$3.65  & 3.81$\pm$0.18 & 1.95$\pm$0.67 & 5.24$\pm$1.62 & 0.09    \\  
10 & 338.194140  & 58.459910   & 6.62$\pm$0.38 & 3.20$\pm$1.18 & -4.16$\pm$1.70 & -9.77 $\pm$1.72 & -5.63$\pm$1.78  & 4.06$\pm$0.08 & 1.91$\pm$0.34 & 9.11$\pm$0.72 & 0.03    \\  
11 & 338.200690  & 58.486168   & 5.89$\pm$0.00 & 6.86$\pm$0.00 & -2.09$\pm$0.00 & -8.01 $\pm$0.00 &  0.26$\pm$0.00  & 4.31$\pm$0.00 & 3.18$\pm$0.00 & 4.64$\pm$0.00 & 4.26    \\  
12 & 338.221150  & 58.468433   & 6.69$\pm$0.60 & 2.98$\pm$1.79 & -3.15$\pm$1.27 & -8.74 $\pm$1.23 & -4.16$\pm$2.63  & 3.98$\pm$0.06 & 1.65$\pm$0.52 & 3.97$\pm$0.41 & 6.83    \\  
13 & 338.238310  & 58.466403   & 5.82$\pm$0.67 & 4.06$\pm$1.28 & -3.70$\pm$1.13 & -9.15 $\pm$1.41 & -3.87$\pm$3.87  & 3.80$\pm$0.13 & 1.98$\pm$0.18 & 8.12$\pm$1.31 & 0.57    \\  
\hline
\multicolumn{11}{c}{YSOs with $H-K~>$ 0.9}\\
\hline
14 & 338.184145  & 58.475136   & 6.15$\pm$0.03 & 3.62$\pm$0.06 & -4.15$\pm$2.11 & -10.27$\pm$1.87 & -4.74$\pm$2.22  & 3.89$\pm$0.03 & 2.04$\pm$0.05 & 3.21$\pm$0.23 & 27.50   \\  
15 & 338.188256  & 58.449883   & 6.34$\pm$0.82 & 3.09$\pm$1.95 & -4.62$\pm$1.18 & -10.42$\pm$1.51 & -3.77$\pm$3.08  & 3.87$\pm$0.12 & 1.53$\pm$0.56 & 4.28$\pm$0.96 & 7.56    \\  
16 & 338.211855  & 58.469717   & 5.10$\pm$0.99 & 5.64$\pm$3.25 & -2.47$\pm$1.40 & -7.63 $\pm$1.96 & -0.40$\pm$2.20  & 3.74$\pm$0.13 & 2.25$\pm$0.89 & 3.87$\pm$1.16 & 0.22    \\  
17 & 338.242297  & 58.478497   & 6.40$\pm$0.63 & 3.28$\pm$1.83 & -4.18$\pm$2.31 & -9.74 $\pm$1.90 & -4.39$\pm$3.49  & 3.89$\pm$0.04 & 1.75$\pm$0.67 & 3.43$\pm$0.39 & 3.52    \\  
18 & 338.245190  & 58.473235   & 4.54$\pm$1.09 & 6.54$\pm$1.60 & -1.70$\pm$1.31 & -6.49 $\pm$0.87 &  0.60$\pm$0.08  & 3.74$\pm$0.12 & 2.73$\pm$0.34 & 4.99$\pm$1.83 & 4.92    \\  
19 & 338.252860  & 58.487277   & 6.35$\pm$0.45 & 3.68$\pm$1.23 & -2.86$\pm$0.74 & -8.97 $\pm$0.86 & -1.82$\pm$1.87  & 4.05$\pm$0.08 & 2.11$\pm$0.58 & 4.07$\pm$1.03 & 3.87    \\
\hline
\end{tabular}
\label{table3}
\end{table}
\end{landscape}
\section {Discussion}
\label{sec_discussion}
\citet{deharveng99} discussed the stellar cluster in the Sh2-138 region using NIR data and compared it with the cluster found in the Sh2-106 region. Furthermore, they suggested that the presence of four massive stars, including
 a probable Herbig Ae/Be star in the central part of the Sh2-138 cluster, has a similar configuration to that found in the Orion Trapezium cluster. We found two among them are $H\alpha$ emission stars in our slit-less spectra.
 In our selected 4$\farcm$6$\times$4$\farcm$6 area in the Sh2-138 region, a detailed analysis of stellar content suggests the presence of a cluster centered on the $IRAS$ source position which mainly contains low-mass stars
 along with a few embedded massive stars. The CM-D analysis ($J/J-H$) has revealed that most of the sources have masses from 0.5--4.0 M$_\odot$ for visual extinction ranged from $\sim$3.0 to $\sim$9.0 mag (see
 Figure~\ref{fig20}). The five radio clumps are also located within the cluster, and only one of them (i.e. clump `A' ), which is estimated to be excited by an O9.5V star (see Table~\ref{table2}), is actually associated with
 at least three sources including the central bright source, as mentioned before. Estimation of the Lyman continuum flux for the remaining four radio clumps reveals that they are associated with sources earlier than B0.5V type, which
 are deeply embedded in the dense cloud without any NIR counterparts. This particular result indicates the formation of young massive stars in the cluster and that the region could be ionized by a small cluster of massive stars.

The bright source associated with the radio clump `A'  was characterized as a Herbig Ae/Be star by \citet{deharveng99}, however, they suggested the presence of non-resolved binary (or multiple) system with the source,
 based on the radio spectral type and NIR data. We characterize this source as a Herbig Be star with the help of the multi-wavelength spectroscopic data ($\sim$0.5--15 $\mu m$). The source is associated with the enhanced
 $H\alpha$ emission and has stellar mass $\sim$ 9M$_\odot$. The dynamical age of the radio clump `A' is estimated to be about 0.16 to 0.54 Myr. In the field of massive star formation, it is often argued whether massive stars
 form before, after, or contemporary with the formation of low-mass stars \citep{tan14}. In this work, the average age of the low-mass stars appears to be higher than the ages of massive stars, indicating the formation of
 low-mass stars prior to the formation of massive stars. This result could be explained by the ``outflow-regulated clump-fed massive star formation'' model of \citet{wang10}, where outflows fragment the filaments and choke
 the mass accretion rate such that massive stars gain masses gradually and form at the end of the cluster formation. High-resolution NH$_{3}$ line observations will be helpful to further investigate this theoretical explanation
 \citep{busquet13}.

\begin{figure}
\centering
\includegraphics[width=0.45\textwidth]{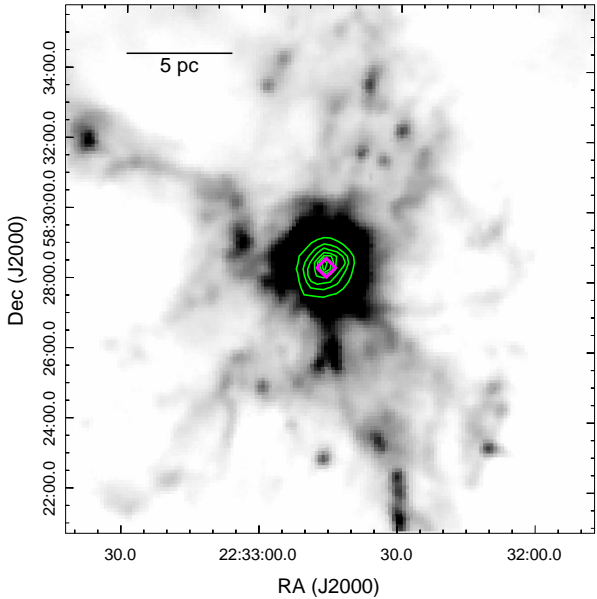}
\caption{20NN surface density contours of YSOs identified in 15$' \times$15$'$ region from UKIDSS catalog are overlaid on the 15$' \times$15$'$ {\it Herschel} 250 $\mu m$ image. The position of the $IRAS$ source is marked
 by a diamond symbol (`$\Diamond$'). The contour levels are 5, 8, 11, 15, 18, 21, and 25 YSOs pc$^{-2}$.}
\label{fig21}
\end{figure}

Our multi-wavelength study is mainly concentrated towards the central 4$\farcm$6$\times$4$\farcm$6 area in the Sh2-138 region. In order to examine the global star formation picture in the Sh2-138 region, we performed analysis
 of stellar content in the 15$' \times$15$'$ area using YSOs identified from UKIDSS-GPS catalog. We generated 20NN density contours of YSOs following the method as discussed in Section~\ref{sec_nn_density}. The 20NN
 surface density contours overlaid on the $Herschel$ 250 $\mu m$ image are shown in Figure~\ref{fig21}. Several prominent parsec-scale filamentary structures can be seen in the $Herschel$ 250 $\mu m$ image (as mentioned in
 Section~\ref{subsecmorp}). The analysis of multi-wavelength data of the Sh2-138 region reveals that the cluster of YSOs harboring massive stars appears to be located at the junction of these filaments. Similar `hub-filament'
 configuration has been reported in other cloud complexes, such as Taurus, Ophiuchus, and Rosette  \citep[e.g.][]{myers09,schneider12}. In Rosette Molecular Cloud, \citet{schneider12} utilized the {\it Herschel} data along
 with the simulations of \citet{dale11}, and suggested that the infrared clusters were located at the junction of filaments or filament mergers. According to \citet{myers09}, a `hub' region should have a typical peak column
 density of $\sim 10^{22}$ cm$^{-2}$ which radiates several parsec-scale filaments associated with column densities of $\sim 10^{21}$ cm$^{-2}$. The filamentary structures identified in the Sh2-138 region appear to follow the
 similar conditions. The mean column density in the hub is $\sim3\times10^{22}$ cm$^{-2}$ and the hub is associated with the peak of 20NN contours of $\sim$50 YSOs pc$^{-2}$ (also see Figure~\ref{fig16}). Also, a cluster of
 YSOs is located with the highest density and the highest temperature region (see Figure~\ref{fig9}). The high temperature of the central part suggests heating of gas from the energetics of massive stars located within the
 cluster. Formation of massive stars at the junction of filaments has also been found recently in the W40 region \citep{mallick13} and IRAS 16148-5011 \citep{mallick15}. The multi-wavelength analysis of the central cluster
 (4$\farcm$6$\times$4$\farcm$6) and the preliminary results in a larger area (15$\arcmin\times$15$\arcmin$) suggest that an isolated cluster, which contains low and massive stars, is being formed at the `hub' of filaments in
 the Sh2-138 region. In future, it will be useful to study the motion of the molecular material along the filaments to further investigate the role of filaments for the formation of YSO cluster.
\section{Conclusions}
We study the physical environment of the Sh2-138 region, a Galactic compact H\,{\sc ii} region, using multi-wavelength observations. We use new optical and NIR photometric and spectroscopic observations, as well as radio
 continuum observations, from Indian observational facilities. Additionally, we explore the region using archival public data covering radio through NIR wavelengths. This work provides a careful look at multi-wavelength
 data from 25 pc to 0.1 pc scale centered on IRAS 22308+5812. {\it Herschel} temperature and column density maps are utilized to examine the physical conditions in the region. Different line ratios are explored to estimate
 the physical properties of the central bright source. We study the various CC-D and CM-D, and the surface density analysis, to investigate the embedded population in the region. The important conclusions of this work are
 as follows:

1. The analysis of optical (0.5-0.9 $\mu m$), NIR (1.0-2.4 $\mu m$), and MIR (5-15 $\mu m$) spectra of the central bright source, located near the $IRAS$ source, reveals that the source is a Herbig Be star. SED modeling of this
 source suggests that the source is young ($\sim$0.03 Myr) and has a mass of $\sim$9 M$_\odot$. In slitless spectra, two sources are identified with strong $H\alpha$ emission and one of them is the central bright source.

2. Using the NIR CC-D, NIR CM-D, and MIR CC-D as tool to distinguish the young stellar sources, we identified a total of 149 YSO candidates in the region and among these 149 young sources, 10 are Class I objects, 54 are
 Class II objects, and the remaining 85 are sources with infrared excess ($H-K>$ 0.9) which could be candidate Class II/Class I YSOs. The $J/J-H$ CM-D analysis shows that the majority of YSOs have masses less than 4 M$_\odot$.
 Our \ks-band luminosity function fits to a slope of ($\alpha$) 0.41$\pm$0.05, typical for young clusters.

3. Previously \citet{deharveng99} found four sources at the central part of the Sh2-138 cluster that are arranged in a similar configuration to that found in the Orion Trapezium cluster. We found that at least three of them
 are younger than 1 Myr and have spectral types earlier than B0, while two among them are $H\alpha$ emission stars.

4. A total of five clumps are identified in the high-resolution 1280 MHz radio continuum map and no optical/NIR counterparts are found for four of them. Assuming a single source is associated with each clump, we estimated
 the spectral types of all the sources to be earlier than B0.5. These results suggest the presence of embedded young massive stars in the Sh2-138 region. Free-free emission SED fitting of the central compact H {\sc ii} clump
 yields an electron density of 2250$\pm$400 cm$^{-3}$, while a lower electron density of 500$\pm$300 cm$^{-3}$ is obtained from the [S {\sc ii}] 6716 to 6731 {\AA} lines' ratio. The reason of this inconsistency is
 unknown. The dynamical age corresponding to the central clump, for an ambient density from 1000 to 10000 cm$^{-3}$, varies between 0.16 to 0.54 Myr.

5. Analysis of {\it Herschel} column density and temperature maps reveals that the region contains a large dust mass of $\sim$3770 M$_\odot$. The NN surface density analysis of YSOs reveals that the YSO cluster is located towards
 the highest column density ($\sim$3$\times$10$^{22}$ cm$^{-2}$) and high temperature ($\sim$35 $K$) regime. The YSO cluster mainly contains low-mass stars as well as a few massive stars. The SED results and the CM-D analysis
 indicate that low-mass stars in the cluster are possibly formed prior to the formation of massive stars. The CO and dust condensation as well as radio continuum emissions are associated with the YSO cluster, where the signature
 of active star formation (i.e. outflow) is evident. Large scale morphology of the region suggests that the YSO cluster is being formed at the junction of {\it Herschel} filaments. 
\section*{Acknowledgments}
We thank the anonymous referee for the useful comments and suggestions which helped to improve the scientific content of the paper. We thank the infrared group of TIFR for their support with TIRSPEC which is extensively used
 for observations of this region. We would like to thank the staff at IAO, Hanle and its remote control station at CREST, Hosakote for their help during the observation runs. We thank the staff at GMRT for their assistance
 during the observations. L.K.D. was supported by the grant CB-2010-01-155142-G3 from the CONACYT (Mexico) during this work. This publication uses data products from the Two Micron All Sky Survey, and the United Kingdom
 Infrared Telescope Infrared Deep Sky Survey. This work is based in part on observations made with the {\it Spitzer} Space Telescope, obtained from the NASA/IPAC Infrared Science Archive, both of which are operated by the
 Jet Propulsion Laboratory, California Institute of Technology under a contract with the National Aeronautics and Space Administration. This research uses the SIMBAD astronomical database service operated at CDS, Strasbourg.
 This publication made use data of 2MASS, which is a joint project of University of Massachusetts and the Infrared Processing and Analysis Centre/California Institute of Technology, funded by the National Aeronautics and Space
 Administration and the National Science Foundation. 

\end{document}